\documentclass[reprint,nofootinbib,amsmath,amssymb,fontenc aps]{revtex4-2}
\usepackage{tabularray}
\usepackage{graphicx}
\usepackage{bm}
\usepackage{placeins}
\usepackage{xspace}
\usepackage{bm,amsmath,amssymb,amsfonts,gensymb,bbold}
\usepackage{comment}
\usepackage{hyperref}
\hypersetup{
    colorlinks=true,
    linkcolor=blue,
    citecolor=blue,
    filecolor=blue,      
    urlcolor=blue,
}
\usepackage{nicefrac}
\usepackage{cleveref}
\usepackage{color}

\newcommand{\eq}[1]{\begin{align} #1 \end{align}}

\newcommand{\tg}{\tilde{g}}
\newcommand{\tp}{\tilde{p}}
\newcommand{\tr}{\tilde{\rho}}
\newcommand{\os}[1]{\textcolor{blue}{\textbf{O.S.: #1}}}

\usepackage[english]{babel}
\usepackage{verbatim}
\usepackage{makecell}

\usepackage{ amssymb }
\begin{document}

\title{Neutron Stars in Causal Scalar-Tensor Theories}

\author{Mark P.~Hertzberg}
\email{mark.hertzberg@tufts.edu}
\author{Oleksandr S.~Stashko}
\email{alexander.stashko@gmail.com}
\affiliation{Institute of Cosmology, Department of Physics and Astronomy, Tufts University, Medford, MA 02155, USA
\looseness=-1}

\begin{abstract}
We study static, spherically symmetric neutron stars in a class of scalar-tensor theories with non-canonical kinetic terms (K-essence) obeying all causality and hyperbolicity conditions. These models have non-trivial dynamics that lead to a type of anti-screening of the scalar. They lead to small corrections in the solar system due to a small coupling, but can lead to large corrections in regimes of high densities, especially neutron stars. We solve the modified Tolman-Oppenheimer-Volkoff equations numerically using realistic equations of state (SLy4, WFF1, MS1, MPA1). For a given central density, we find that two distinct configurations may exist, forming two separate branches of solutions. We find that above a certain critical central density solutions with the correct asymptotic behavior at spatial infinity cannot be obtained. We obtain precise predictions for the mass-radius relation for neutron stars for different values of the parameters in the model and we compare to data. 
\end{abstract}

\maketitle

\section{Introduction}
General Relativity (GR) has been extensively tested across a wide range of physical regimes, from precision experiments in the Solar System to strong-field phenomena such as black hole imaging \cite{2019ApJ...875L...1E} and gravitational wave detections from compact binary mergers \cite{PhysRevLett.116.061102,PhysRevLett.119.161101}. (For a review, see Refs.~\cite{Clifton:2011jh, Will_2014}.)

A fundamental task driving much research (such as observations of black holes) is to test GR against other competing theories. However, it is rather difficult to build a competing theory. This is because GR is the {\em unique} theory of massless spin 2 particles. In order to go beyond GR one must introduce new degrees of freedom, namely new light scalars that can provide a type of fifth force. There are tight constraints on such new scalar forces from solar system tests. However, in this work we will focus on a space of models that was introduced in Ref.~\cite{Hertzberg_2023}. These models are viable from solar system tests as they have a type of anti-screening; the new force is negligible in weak fields, such as the solar system, due to a small coupling, but can be large in strong fields, such as the interior of neutron stars. This space of models is strongly motivated from the fact that a kind of anti-screening is allowed from the point of view of fundamental physics as it maintains subluminality of signals (while screening does the opposite and fails this criteria).


Combining GR with a scalar is often referred to as a ``scalar-tensor" theory. Various versions have been widely studied in the literature \cite{Fujii:2003pa, Faraoni:2004pi, Berti_2015}. However, the very well motivated anti-screened form that we focus on here has not been fully explored. This form results in modified dynamics in the strong-field regime. Neutron stars, with their high compactness and extreme central densities, provide an ideal environment \cite{Yunes_2022, MUSES:2023hyz} to test such deviations. They allow us to probe gravity in conditions far beyond what can be realized in laboratory experiments or Solar System observations.

The structure of neutron stars is affected by the equation of state (EoS) of the nuclear matter. Although nuclear theory and experimental data constrain the EoS near saturation density \cite{Lattimer_2016, MUSES:2023hyz} for  $\rho\lesssim 2\rho_{sat}\simeq 5.4\cdot 10^{14}$ $\textrm{g/cm}^3$, the densities in NS cores can be several times higher, where the behavior of matter becomes highly uncertain. Different theoretical models predict a wide range of high-density EoSs, leading to significant variation in predicted macroscopic properties such as mass, radius, and tidal deformability.
This uncertainty complicates the interpretation of neutron star observations, since both dense matter microphysics and modifications to gravity can influence the same observables. In some cases, EoSs that are excluded under GR may become consistent with data in alternative gravity theories due to modified stellar structure. On the other hand, distinct features arising from modified gravity can produce observable signatures that cannot be mimicked by EoS variation alone. The recent X-ray observations from the NICER mission and gravitational wave detections by LIGO/VIRGO/KARGA put tight constraints on the allowed range of NS parameters  \cite{Miller_2019,Miller_2021,PhysRevLett.121.161101,Abbott_2020,Abbott_2020_1}.

The structure and observable properties of neutron stars have been extensively studied in a variety of modified gravity theories, including scalar-tensor theories \cite{2015CQGra..32n5008S,PhysRevLett.70.2220,Pani_2014,Brown_2023,2021JCAP...10..022B, Cisterna:2016vdx,Maselli_2016,Kase_2020,Cisterna:2015yla,Boumaza:2023wuc}, vector-tensor theories \cite{Ji:2024aeg,Kase_2020P}, higher derivative pure gravity corrections \cite{Liu:2024wvw,Saavedra:2024fzy,Doneva:2023kkz}, etc.

In this paper, we consider static, spherically symmetric neutron stars in this causal class of scalar-tensor theories with non-canonical kinetic terms. These are Lagrangians with some non-trivial function of the kinetic term; which is perfectly allowed by Lorentz symmetry. In the context of cosmological models, related ideas have been proposed to describe inflation and late-time acceleration \cite{Armendariz-Picon:1999hyi,Armendariz-Picon:2000ulo} and in that context are often referred to as K-essence.
Work on using non-canonical kinetic terms in the context of compact objects, includes Refs.~\cite{Cayuso:2024ppe,Brax_2013,Brax_2016,Brax:2021wcv}. However, much of that work involves some kind of screening mechanism. These tend to violate the condition of subluminality and are considered impossible to possess a UV completion. Instead, our work here is on anti-screened models, which could potentially have a UV completion.

The paper is structured as follows. Section~\ref{sec:Basic} presents the theoretical framework and the specific form of the space of models considered. In Section~\ref{sec:NeutronStars}, we derive the modified Tolman–Oppenheimer–Volkoff equations for static, spherically symmetric configurations and discuss the corresponding asymptotic behavior. Section~\ref{sec:NI} outlines the numerical methods and the implementation of the equations of state. The results are presented in Section~\ref{sec:results}. We conclude with a summary and outlook in Section~\ref{sec:conclusions}.

\section{Basic Theory}
\label{sec:Basic}
We consider the specific scalar-tensor theory with the following action in the Einstein frame (signature + - - -) \footnote{We work in units where  $c=h=1$ and $M_p^2=1/G$}
\eq{\label{eq:action}
S=\int d^4x\sqrt{|g|}\left(\frac{R}{16\pi G}+K(X)\right) + S_M(\psi_i,\tg_{\mu\nu}),
}
where $R$ is the scalar curvature constructed from the metric $g_{\mu\nu}$, $S_M$ denotes the action of the  matter fields $\psi_i$  coupled to the Einstein-frame metric and scalar field $\phi$ via via the Jordan-frame metric   $\tilde{g}_{\mu\nu} = f(\phi) g_{\mu\nu}$, $K(X)$ is the $K$-essence pure kinetic term with
\eq{
X=\frac{1}{2}g^{\mu\nu}\partial_{\mu}\phi\partial_{\nu}\phi.
}
For nomenclature, we will refer to this as the ``Einstein frame" in which matter interacts both gravitationally and with the scalar field, and the
the ``Jordan frame" as the frame in which the matter fields interact only gravitationally, unaffected by the scalar field.  Throughout this work, quantities defined in the Jordan frame are denoted with a tilde, e.g $\tilde{x}$.

Variation of  (\ref{eq:action}) with respect to the metric tensor $g_{\mu\nu}$ yields the Einstein equations
\eq{
G_{\mu\nu}=8\pi G\left(T_{\mu\nu}^{(SF)}+T_{\mu\nu}^{(M)}\right),
}
where the energy-momentum tensor of the scalar field is given by 
\eq{
T_{\mu\nu}^{(SF)} = K'_X\partial_{\mu}\phi\partial_{\nu}\phi-g_{\mu\nu}K,~K'_X=\frac{dK}{dX},
}
and $T_{\mu\nu}^{(M)}$ denotes the energy-momentum tensor of the matter fields defined in the standard way using both metrics
\eq{
T_{\mu\nu}^{(M)}=\frac{2}{\sqrt{|g|}}\frac{\delta S_M(\psi_i,\tg_{\mu\nu})}{\delta g^{\mu\nu}},~\tilde{T}_{\mu\nu}^{(M)}=\frac{2}{\sqrt{|\tilde{g}|}}\frac{\delta S_M(\psi_i,\tg_{\mu\nu})}{\delta \tilde{g}^{\mu\nu}}
,}
related through conformal coupling as
\eq{
T_{\mu\nu}^{(M)}=f(\phi) \tilde{T}_{\mu\nu}^{(M)}.
}

Variation of  (\ref{eq:action}) with respect to the scalar field $\phi$ leads to the following equation of motion

\eq{
\partial_{\mu}\left(K'_X\sqrt{|g|}g^{\mu\nu}\partial_{\mu}\phi\right)= -\frac{\sqrt{|g|}f'}{2f}T^{(M)},
}
where $T^{(M)}=g^{\mu\nu}T^{(M)}_{\mu\nu}$ is a trace.

Using the Bianchi identities, $\nabla_{\mu} G^{\mu\nu}$, one obtains the equation of motion for the matter fields in the Einstein frame
\eq{
\nabla_{\mu} T^{\mu\nu}{}^{(M)}=-\nabla_{\mu} T^{\mu\nu}{}^{(SF)}.
}
Due to the conformal coupling, the motion of a massive particle in the Einstein frame is non-geodesic, as it experiences an additional force $F^{\mu}$ that is orthogonal to the particle’s four-velocity.
\eq{
\frac{d^2 x^{\mu}}{d\tau^2}+\Gamma^{\mu}_{\alpha\beta}\frac{d x^{\alpha}}{d\tau}\frac{d x^{\beta}}{d\tau}=F^\mu,
}
where 
\eq{
F^\mu=\frac{f'}{2f}\left(g^{\sigma \mu}-\frac{d x^{\sigma}}{d\tau}\frac{d x^{\mu}}{d\tau}\right)\partial_{\sigma}\phi, ~g_{\sigma\mu}F^{\mu}\frac{d x^{\sigma}}{d\tau}=0.
}
and $\tau$ is a canonical parameter.

We consider a class  of $K$-essence models where the kinetic function $K(X)$  satisfies standard causality and hyperbolicity conditions \cite{PhysRev.182.1400,Rendall_2006,Hertzberg_2023}
\eq{
K'_X\geq0,~~K''_{XX}\geq0,~~K'_X+2XK''_{XX}>0.
}
In particular, we restrict ourselves to the following space of kinetic functions $K(X)$ \cite{Hertzberg_2023} as it obeys all these causality and hyperbolicity conditions
\eq{
K(X)=-\frac{\mu}{\gamma}\left[\left(1-\frac{X}{\mu}\right)^\gamma-1\right],\,\,\,~\frac{1}{2}<\gamma<1.
}
The form here has $K\approx X$ for low densities, where it becomes canonical. But enters a new scaling regime $K\propto (-X)^\gamma$ for high densities (note that for static fields $X$ is negative in our signature). With $1/2<\gamma<1$ this new scaling regime has some interesting properties. In particular, the causality bound, which demands subluminality of signals and $\gamma<1$ (as the speed of signals in the high density regime can be shown to be $c_s=\sqrt{2\gamma-1}$ \cite{Hertzberg_2023}),
leads to a type of ``anti-screening", in which, contrary to conventional screening mechanisms \cite{Koyama_2016} (which suffer from causality problems), the scalar-mediated force is amplified in high-density regions \cite{Hertzberg_2023}.

To describe compact objects such as neutron stars, we model the matter sector as a ideal isotropic fluid, with stress-energy tensor
\eq{\label{eq:Tmunu_ein_or}
\tilde{T}^{(M)}_{\mu\nu}=(\tilde{p}+\tilde{\rho})\tilde{u}_{\mu}\tilde{u}_{\nu}-\tilde{p}\tilde{g}_{\mu\nu},
}
where $\tilde{\rho}$,  $\tilde{P}$, and $\tilde{u}_{\mu}$  are energy density, pressure and fluid's four-velocity, respectively.

The relations between physical quantities in the two frames are
\eq{
u_{\mu}=\frac{1}{\sqrt{f}}\tilde{u}_{\mu},~\rho=f^2\tilde{\rho},~p=f^2\tilde{p},
 }
allow us to rewrite  the matter energy-momentum tensor $(\ref{eq:Tmunu_ein_or}) $  in the Einstein frame as 
\eq{
T^{(M)}_{\mu\nu}=f^2(\phi)[(\tilde{p}+\tilde{\rho})u_{\mu} u_{\nu}-\tilde{p}g_{\mu\nu}].
}
\section{Static neutron stars}
\label{sec:NeutronStars}
We consider a static, spherically symmetric spacetime with the following line element in the Einstein frame
\eq{
ds^2=e^{2\alpha(r)}dt^2-e^{2\beta(r)}dr^2-r^2 d\Omega^2,
}
where $d\Omega^2=d\theta^2+\sin^2\theta d\phi^2$. 

For convenience, we introduce the  mass function,
\eq{
M(r)=\frac{r}{2G}\left(1-e^{-2\beta(r)}\right),
}
which coincides with the total ADM mass of the configuration at spatial infinity, $M_{ADM}=\lim\limits_{r\to{\infty}}M(r)$.

Then the non-trivial  equations of motion can be written in the the following form
\eq{
\label{eq:EoM1}
\frac{dM}{dr}=4\pi G r^2 \left(f^2 \tilde{\rho}-K(X)\right),
}

\eq{
\label{eq:EoM2}
\frac{d\alpha}{dr}=\frac{GM+ 4 \pi G  r^3
   \left(K(X)+f^2\tilde{p}\right)}
   {r (r-2G M)}+4 \pi G  r \phi'^2
   K'_X,
}

\eq{\label{eq:EoM3}
\frac{d}{dr}\left[r^2 e^{\alpha-\beta} K'_X\phi'\right]=\frac{1}{4}r^2e^{\alpha+\beta}(\tilde{\rho}-3\tilde{p})\frac{d}{d\phi}f^2(\phi), }

\eq{\label{eq:EoM4}
\frac{d\tilde{p}}{dr}=-(\tilde{p}+\tilde{\rho})\left[\alpha'+\frac{d}{dr}\ln \sqrt{f(\phi)}\right],
}
which, with an equation of state (EoS) $\tilde{\rho}=\tilde{\rho}(\tilde{p})$ form a modified Tolman-Oppenheimer-Volkoff system.  

We are interested in the regular solutions at  $r\to{0}$, which requires the following asymptotic behavior

\eq{\label{eq:asymp_at_center}
&\alpha(r)\sim \alpha_c,~~M(r)\sim \frac{4}{3}\pi f^2(\phi_c)\tr_c r^3,\nonumber\\
&\phi(r)\simeq \phi_c, ~~ \tp(r) \simeq \tp_c,
}
where the subscript $c$ denotes central values of the corresponding quantities.

Outside the star, the pressure and energy density vanish, 
 $\tilde{p}=\tilde{\rho}\equiv 0$, the system reduces to a pure vacuum $K$-essence configuration. In general, exact  analytical solutions for this system are not known, except the case with $ p = 1$, where  the exterior solution reduces to the FJNW solution \cite{fisher1999scalarmesostaticfieldregard,PhysRevLett.20.878}.

 At spatial infinity $r\to\infty$,  the solutions can be expanded in powers of $1/r$.  The leading-order terms of the  asymptotic behavior of the metric and scalar field are

\eq{\label{eq:asympt_at_infinity}
e^{2\alpha(r)}= 1-\frac{2 G M}{r}+\frac{4  \pi G^3  M
   Q^2}{3 r^3}+\mathcal{O}\left(\frac{1}{r^4}\right),
}
\eq{\label{eq:asympt_at_infinity_1}
M(r)= M-\frac{2 \pi
    GQ^2}{r}+\frac{2 \pi G^2  M Q^2}{r^2}+\mathcal{O}\left(\frac{1}{r^3}\right),
    }
\eq{\label{eq:asympt_at_infinity_3}
\phi = \frac{GQ}{r}+\frac{G^2M
   Q}{r^2}+ \frac{2G^3 \left(2 M^2 Q-\pi 
   Q^3\right)}{3 r^3}+\mathcal{O}\left(\frac{1}{r^4}\right),
}
where $M$ is the total ADM mass of the configuration, and $Q$ is the scalar charge that characterizes the strength of  the scalar field, respectively.

We should to note  that  the first terms of the asymptotic expansion at the center of star and at spatial infinity coincide with similar expansions  for the canonical massless scalar field as $K(X)\sim X$ for $r\to{\infty}$ or $r\to{0}$. 

However, the $K$-essence effects became significant at intermediate radii (see Fig. \ref{fig:K-essence}).

The scalar field equation can be integrated, yielding a value for a type of enclosed $Q_{\rm enc}$ at any $r$ of
\eq{
GQ_{\rm enc}(r)=-r^2 e^{\alpha(r)}\sqrt{1-2GM(r)/r}\, K'_X\,\phi'(r),
\label{eq:Qphi}}
with the asymptotic value 
\eq{
Q=Q_{\rm enc}(r\to\infty)
}
being the scalar charge. We note that this can be written in terms of the integral of the trace of the matter energy-momentum tensor as
\eq{
GQ_{\rm enc}(r)=-\int_0^rdy\,y^2 e^{\alpha(y)+\beta(y)}{f'(y)\over 2f(y)}\,T^{(M)}(y),
\label{eq:QT}
}
(here $y$ is just a dummy variable of integration). As we discuss below, we will focus on a form of $f$ whereby $f'/f=\beta_1$ is a positive constant; so this factor is very simple. This means that the charge is proportional to the integral of $T^{(M)}$. Outside the star we have $T^{(M)}=0$, so the total charge $Q$ can be expressed as this volume integral over the interior of the star. Nevertheless, we find it more convenient to numerically determine the charge $Q$ from the derivative of the asymptotic value of $\phi$.


Our physical metric $\tilde{g}_{\mu\nu}$, defined in the Jordan frame can be written as \eq{
d\tilde{s}^2=e^{2\tilde{\alpha}(\tilde{r})}d\tilde{t}^2-e^{2\tilde{\beta}(\tilde{r})}d\tilde{r}^2-\tilde{r}^2 d\Omega^2,
}
and they can be related to the Einstein's frame counterparts as
\eq{
\tilde{t}=t,~\tilde{r}=\sqrt{f}\,r,~e^{2\tilde{\alpha}}=f(\phi)e^{2\alpha},~e^{2\tilde{\beta}}=fe^{2\beta}\left(\frac{d\tilde{r}}{dr}\right)^{-2},
}

To proceed further, we must specify the form of the conformal coupling $f(\phi)$.
In this work, we use a simple exponential form
\eq{
f(\phi)=e^{\beta_1 \phi}
}
Solar-system experiments impose strong constraints on  possible deviations from general relativity via  PPN formalism. At the $1\sigma$ level, the PPN parameters are constrained as \cite{Bertotti:2003rm,2004PhRvL..93z1101W}
\begin{eqnarray}
&&\gamma_{\textrm{PPN}}-1=(2.1\pm2.3)\cdot10^{-5},\\
&&\beta_{\textrm{PPN}}-1=(-4.5\pm5.6)\cdot 10^{-5}.
\end{eqnarray}
 which lead to an upper bound for the scalar coupling parameter $\beta_1$ as 
 
\eq{\beta_1\lesssim0.024/M_p.}

Using the asymptotic expansions (\ref{eq:asympt_at_infinity},\,\ref{eq:asympt_at_infinity_1},\,\ref{eq:asympt_at_infinity_3}), we  can obtain relations between the ADM mass $\tilde{M}_{ADM}$ in the Jordan and Einstein frames
\eq{
\tilde{M}_{ADM}=M+ \frac{1}{2}\beta_1 Q,
}
while the active gravitational mass $\tilde{M}$ defined by the asymptotic behavior of $e^{2\tilde{\alpha}}$ and has the following form
\eq{
\tilde{M}=M- \frac{1}{2}\beta_1 Q.
}

\section{Numerical implementation}
\label{sec:NI}
For numerical computations, we will make reference to the following characteristic mass density:
\eq{
\rho_0\equiv 10^{15}\,\textrm{g}/\textrm{cm}^3
}
Using this, we introduce the following characteristic scales
\begin{eqnarray}
&&\mu_0 =  c^2\rho_0\simeq 
(0.256\,\textrm{GeV})^4\\
&&r_0=c /\sqrt{G\rho_0}\simeq36.7\,\textrm{km}\\
&&M_0=c^3 /\sqrt{G^3\rho_0} \simeq 24.8M_\odot
\end{eqnarray}
We then define dimensionless variables $\{\mu',\rho',p',r',M'\}$
through
\eq{
\mu=\mu_0 \mu',\,\rho=\mu_0\rho',\,p=\mu_0 p',\,r= r_0r',\,M= M_0M'.}
Also, we use the Planck mass $M_p$ to obtain a dimensionless value of scalar field and coupling constant
\eq{
\phi=M_p\varphi,\,\beta_1=\beta_1'/M_p.}

To find solutions of equations (\ref{eq:EoM1})--(\ref{eq:EoM4}) that satisfy the regularity conditions at the center of star (\ref{eq:asymp_at_center}) and the required asymptotic behavior at spatial infinity, we employ a shooting method.

We specify the central pressure $\tilde{p}_c$ and initially set $\alpha_c = 0$. The central value of the scalar field, $\phi_c$, is treated as the shooting parameter. The system (\ref{eq:EoM1})--(\ref{eq:EoM4}) is then integrated outward from the star's center  until the pressure vanishes at some radius $r' = R_s$, defining the stellar surface, i.e., $\tilde{p}'(R_s) = 0$.
The integration is then continued into the exterior vacuum region, where $\tilde{p} = \tilde{\rho} = 0$, up to a sufficiently large radius $r' = r_{\infty}' \gg R_s$. In our calculations,  we set $r_{\infty}' = 10^5$. By varying $\phi_c$, we search for solutions that satisfy the correct asymptotic behavior at spatial infinity. Specifically, we require that at $r = r_\infty$, the scalar field decays to $|\phi_\infty| = \min(10^{-6}, 10^{-5} |\phi_s|)$, where $\phi_s$ is the value of the scalar field at the stellar surface. We verified that small changes of $|\phi_\infty|$ do not significantly affect the results. Due to the shift symmetry of the metric function $\alpha(r)$, the corrected central value can be restored by setting $\alpha_c = -\alpha(r_\infty)$.

Since the true EoS for nuclear matter is currently unknown, we rely on numerical calculations intended to provide a realistic description of nuclear matter. 
We use four popular barotropic equations of state \cite{PhysRevLett.121.161101}. Specifically, we consider WFF1 \cite{PhysRevC.38.1010}, a soft EoS derived from variational calculations using realistic two- and three-body nuclear forces; SLy4 \cite{Douchin_2001}, based on the Skyrme Lyon interactions,
MS1 \cite{M_ller_1996}, based on the relativistic mean-field model and MPA1 \cite{1987PhLB..199..469M} which is based on Brueckner–Hartree–Fock calculations for nuclear matter

\begin{figure}[t]
    \includegraphics[width=\columnwidth]{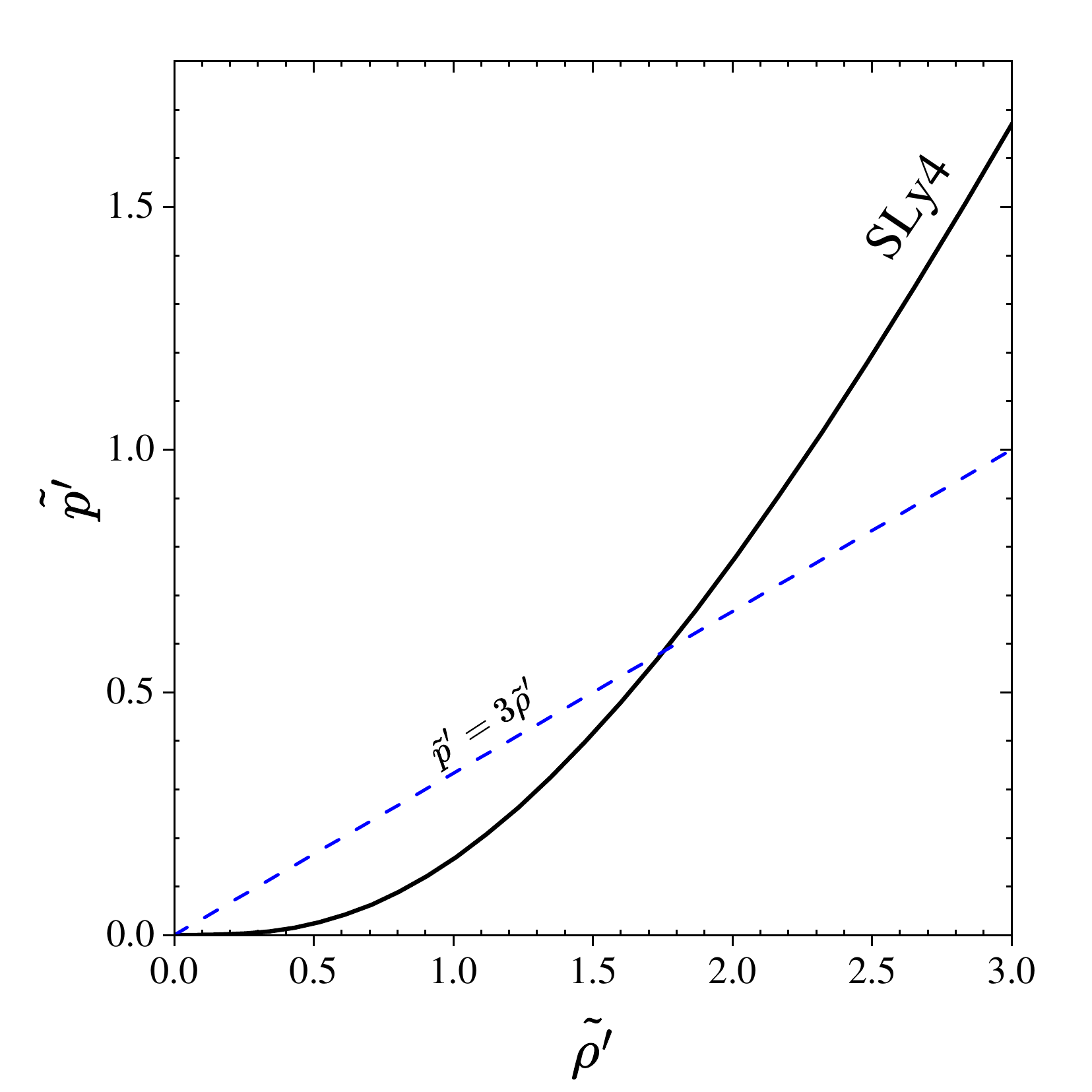}%
    \caption{An example of the SLy4 EoS \cite{Douchin_2001}; The blue dashed  line corresponds to the conformal limit ($\tilde{p}=\tilde{\rho}/3$). One can see that $\tilde{T}^{(M)}=\tilde{\rho}-3\tilde{p}$ is positive for low densities (relevant to outer part of neutron star), while it is negative for high densities (relevant to inner part of neutron star).
    }
    \label{fig:EoS}
\end{figure}

The original EoSs  are provided in tabulated form; an example of this is shown in Figure \ref{fig:EoS} for SLy4. We use this with an interpolation. Alternatively, analytical representations of the corresponding EoSs can also be employed \cite{Haensel_2004, gungor2011analyticalrepresentationequationsstate}, or piecewise polytropic approximations can be used \cite{Read_2009}.

The tabulated EoS are defined with linear interpolation over the interval  $\tilde{p}\in (\tilde{p}_{min},\tilde{p}_{max})$. Additionally, the upper limit $\tilde{p}_{max}$ may be restricted by causality, which requires the speed of sound to satisfy $c_s^2=\partial \tilde{p}/\partial \tilde{\rho} \leq 1$. 

We note that, as seen in Figure \ref{fig:EoS}, the EoS obeys $p<\rho/3$ (i.e., $T^{(M)}=\rho-3p>0$) for low densities, but has $p>\rho/3$ (i.e., $T^{(M)}=\rho-3p<0$) for high densities. Recall from Eq.~(\ref{eq:QT}) the enclosed scalar charge is related to the negative of the integrated $T^{(M)}$. This means that the enclosed charge $Q_{\rm enc}(r)$ is positive for small $r$, where the densities are highest. In this regime it follows from Eq.~(\ref{eq:Qphi}) that $\phi'<0$, and so the corresponding scalar force on a test particle is {\em repulsive}. On the other hand, we will find that the total integrated charge $Q$ is negative (since the outer parts of the star have $T^{(M)}>0$ which over compensates), resulting in $\phi'>0$ and an {\em attractive} scalar force at large distances.

\section{Predictions for neutron stars}
\label{sec:results}
We solve the modified TOV equations(\ref{eq:EoM1})–(\ref{eq:EoM4}) to obtain static equilibrium configurations of neutron stars as functions of the model parameters $(\mu', \gamma)$ and the central values of the fluid pressure and scalar field, $(\tilde{p}'_c, \varphi_c)$, respectively.

\begin{figure*}
    \includegraphics[width=0.49\linewidth]{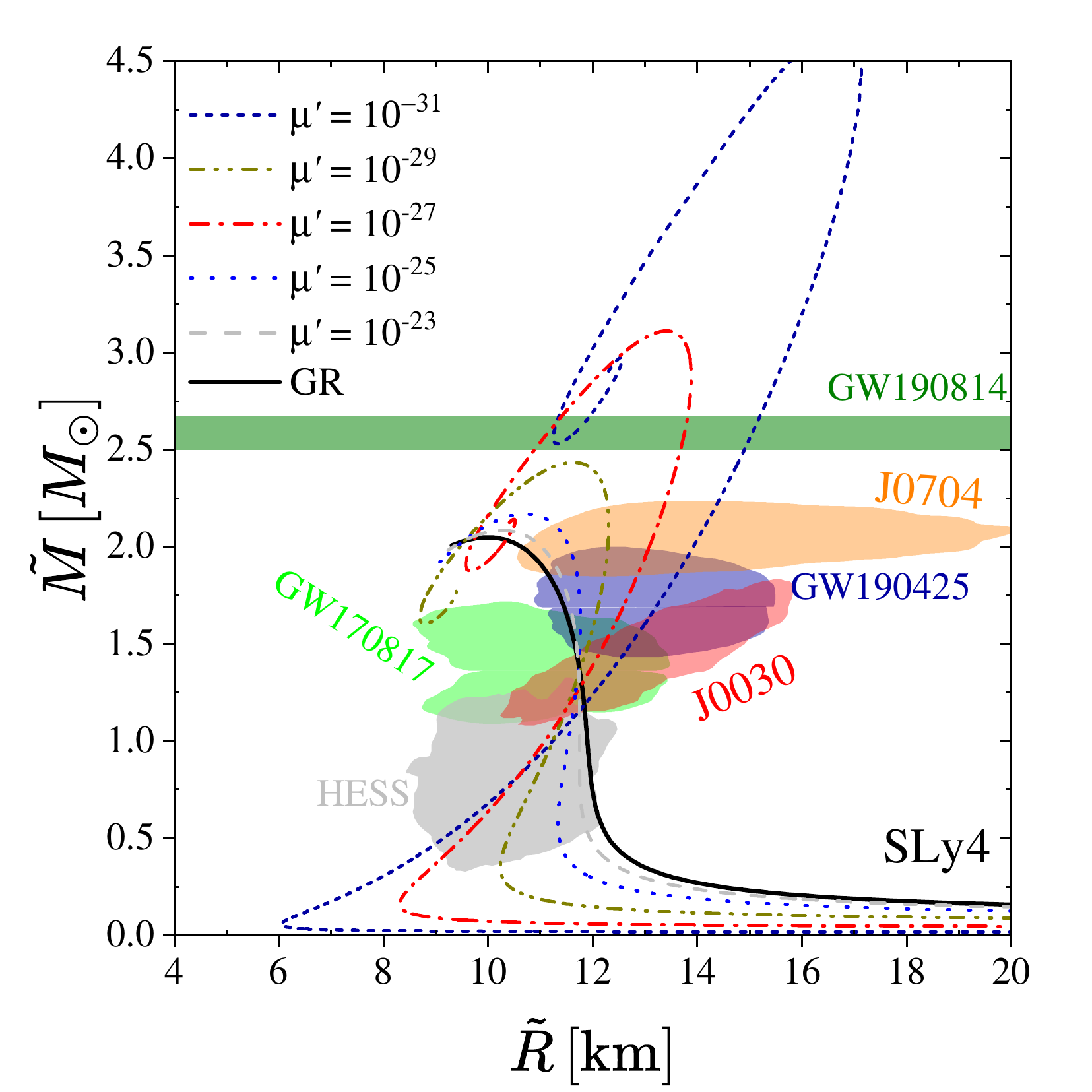}%
     \includegraphics[width=0.49\linewidth]{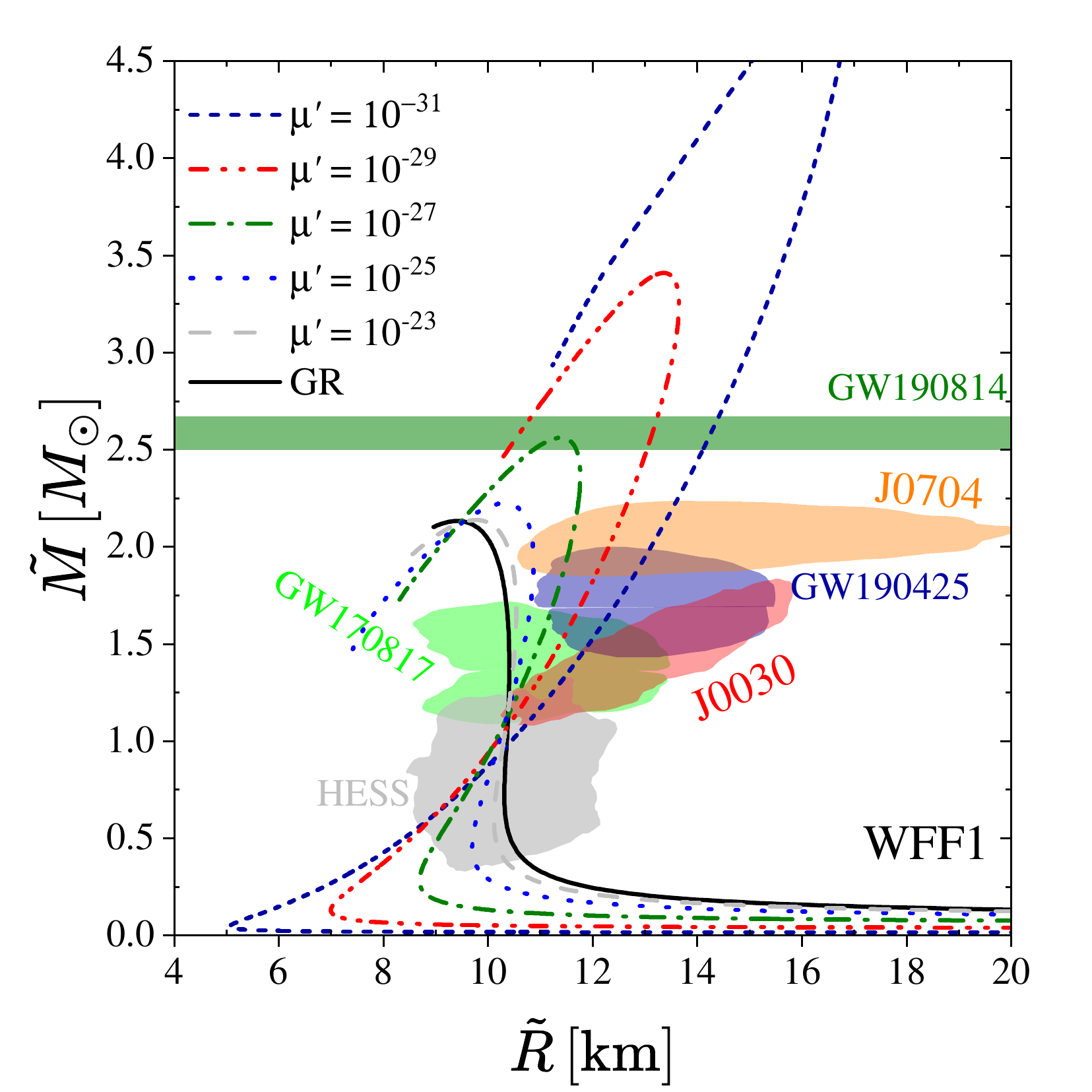}%
     \\
      \includegraphics[width=0.49\linewidth]{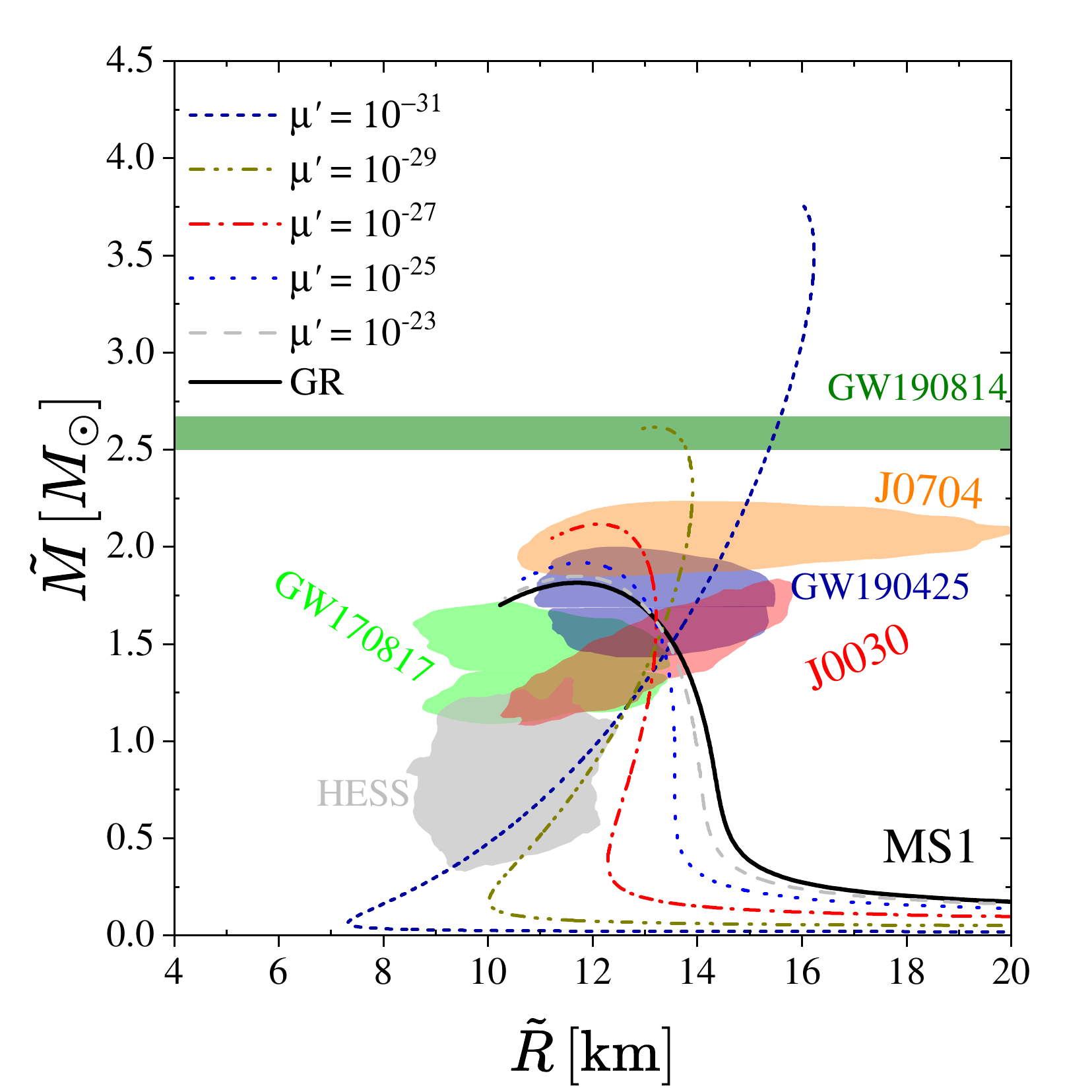}%
      \includegraphics[width=0.49\linewidth]{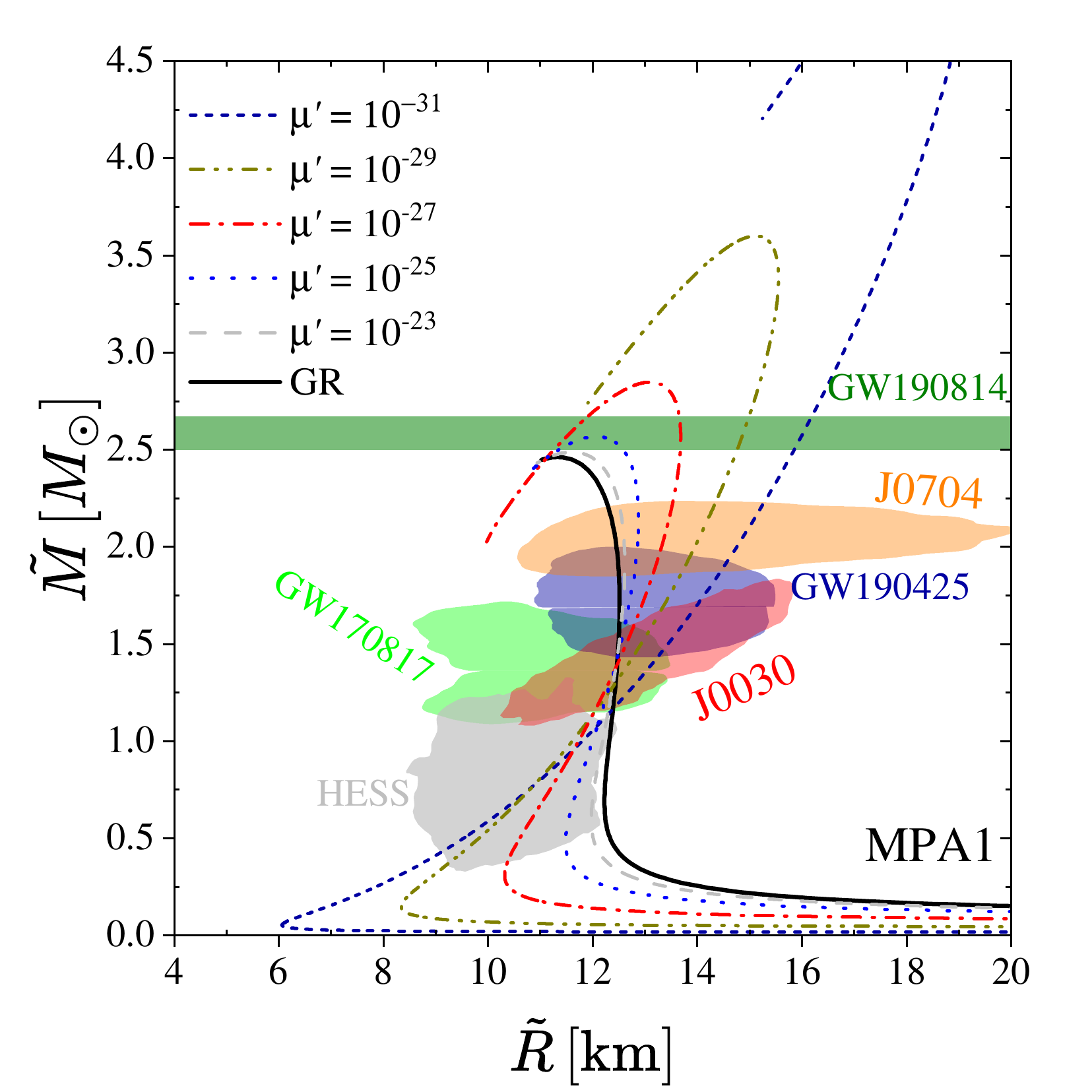}
    \caption{ Mass-radius diagram for neutron stars for different realistic equations of state with $\gamma=5/6$, $\beta_1'=0.024$ for different values of $\mu$. The colored domains correspond to the  recent observational mass-radius  constraints by the gravitational wave events GW170817 \cite{PhysRevLett.119.161101,PhysRevLett.121.161101}, GW190425\cite{Abbott_2020}, and GW190814 \cite{Abbott_2020_1} at $90\%$ credible level,   pulsars  PSRJ0030-0451\cite{Miller_2019}, PSRJ0740-6620\cite{Miller_2021} and HESS J1731-347 object \cite{2022NatAs...6.1444D}, at $95\%$ credible level, correspondingly. }
    \label{fig:MR1}
\end{figure*}

\begin{figure*}
    \includegraphics[width=0.49\linewidth]{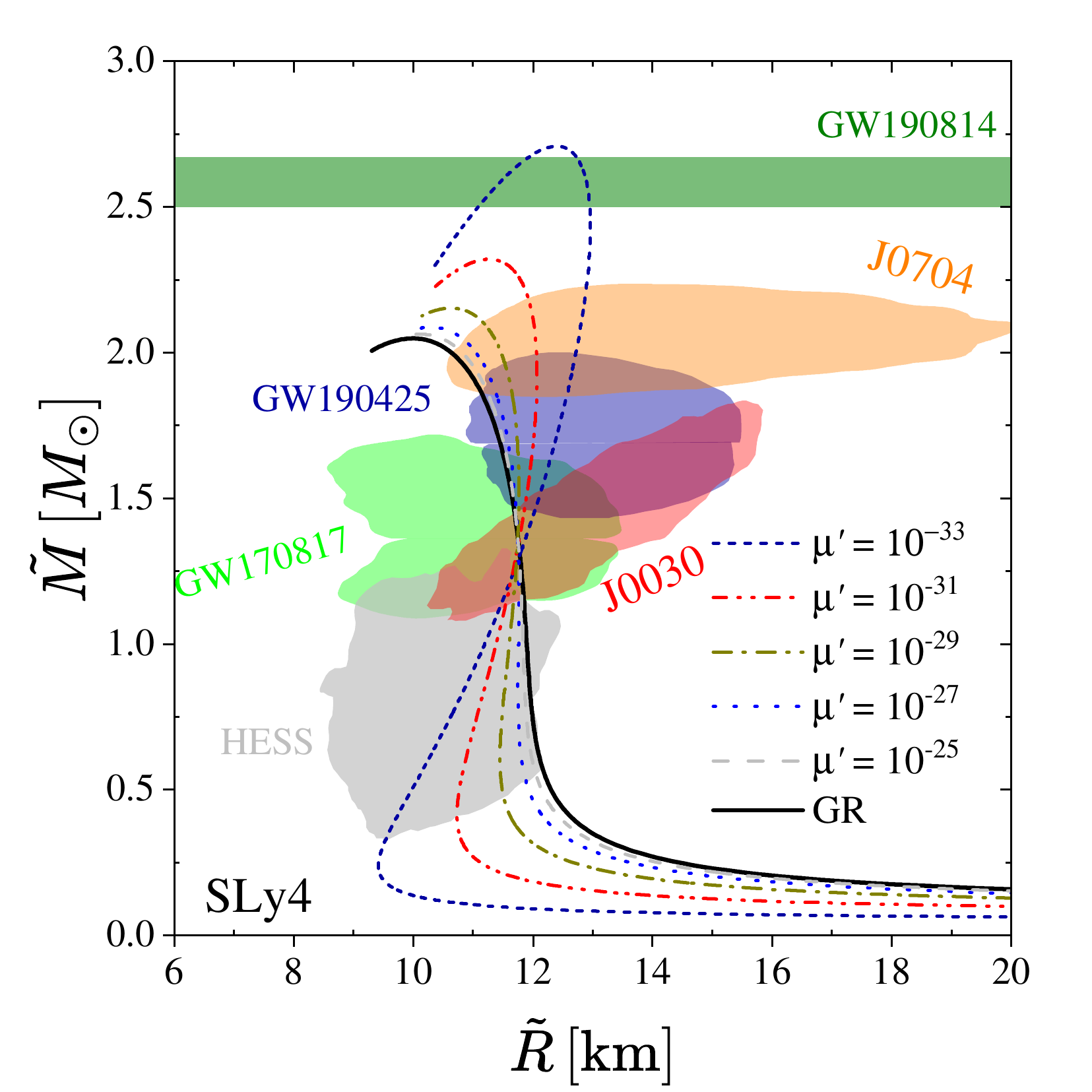}%
     \includegraphics[width=0.49\linewidth]{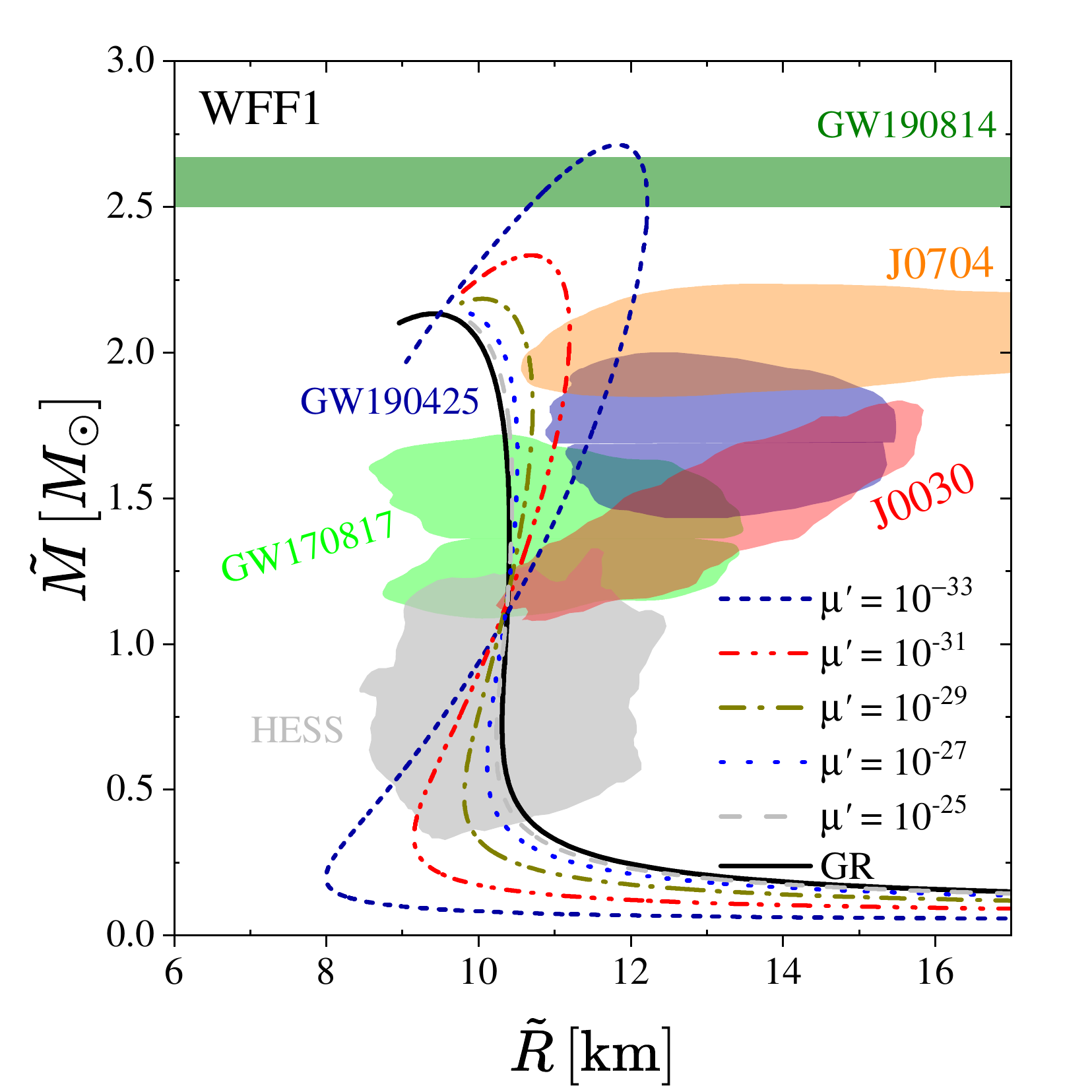}
      \includegraphics[width=0.49\linewidth]{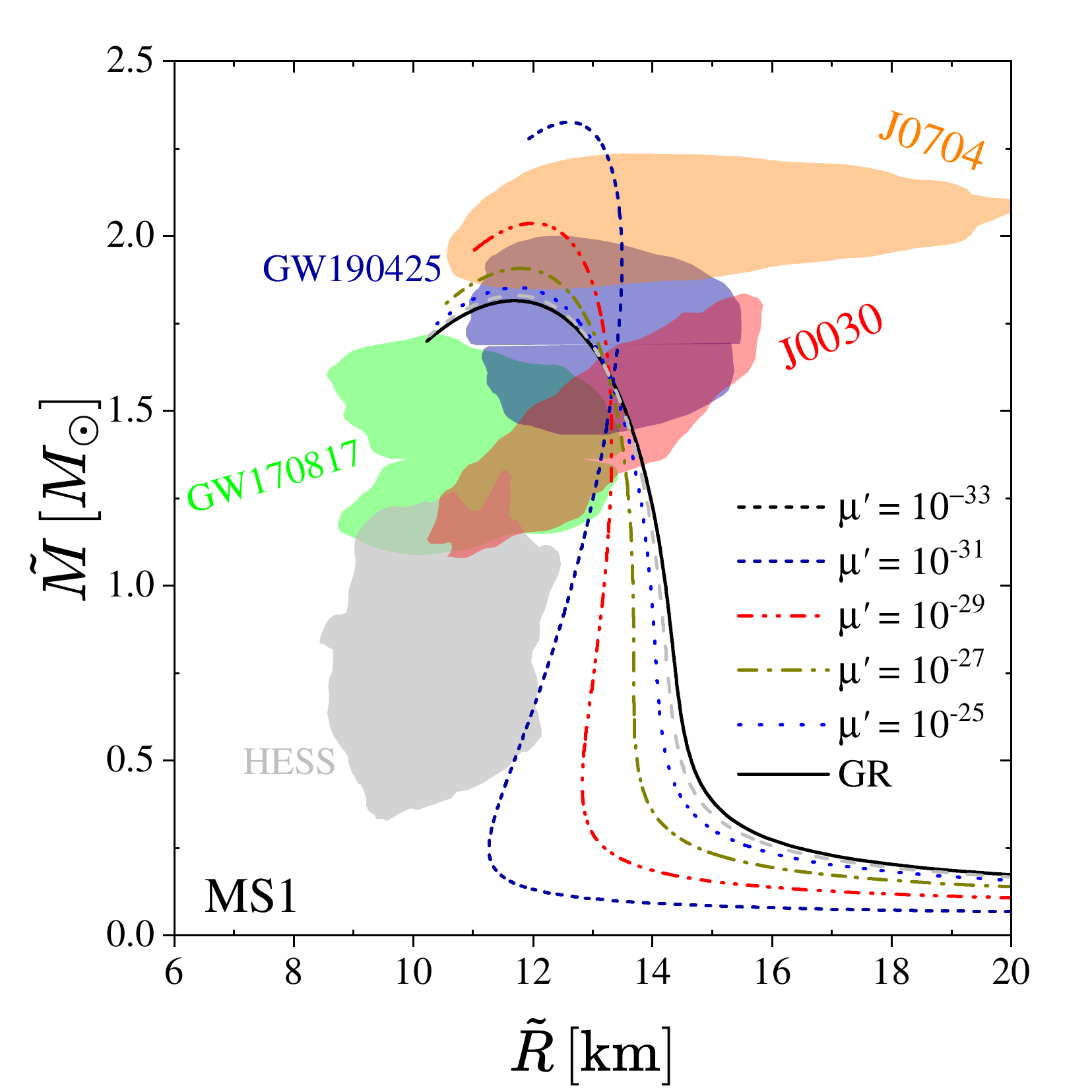}%
      \includegraphics[width=0.49\linewidth]{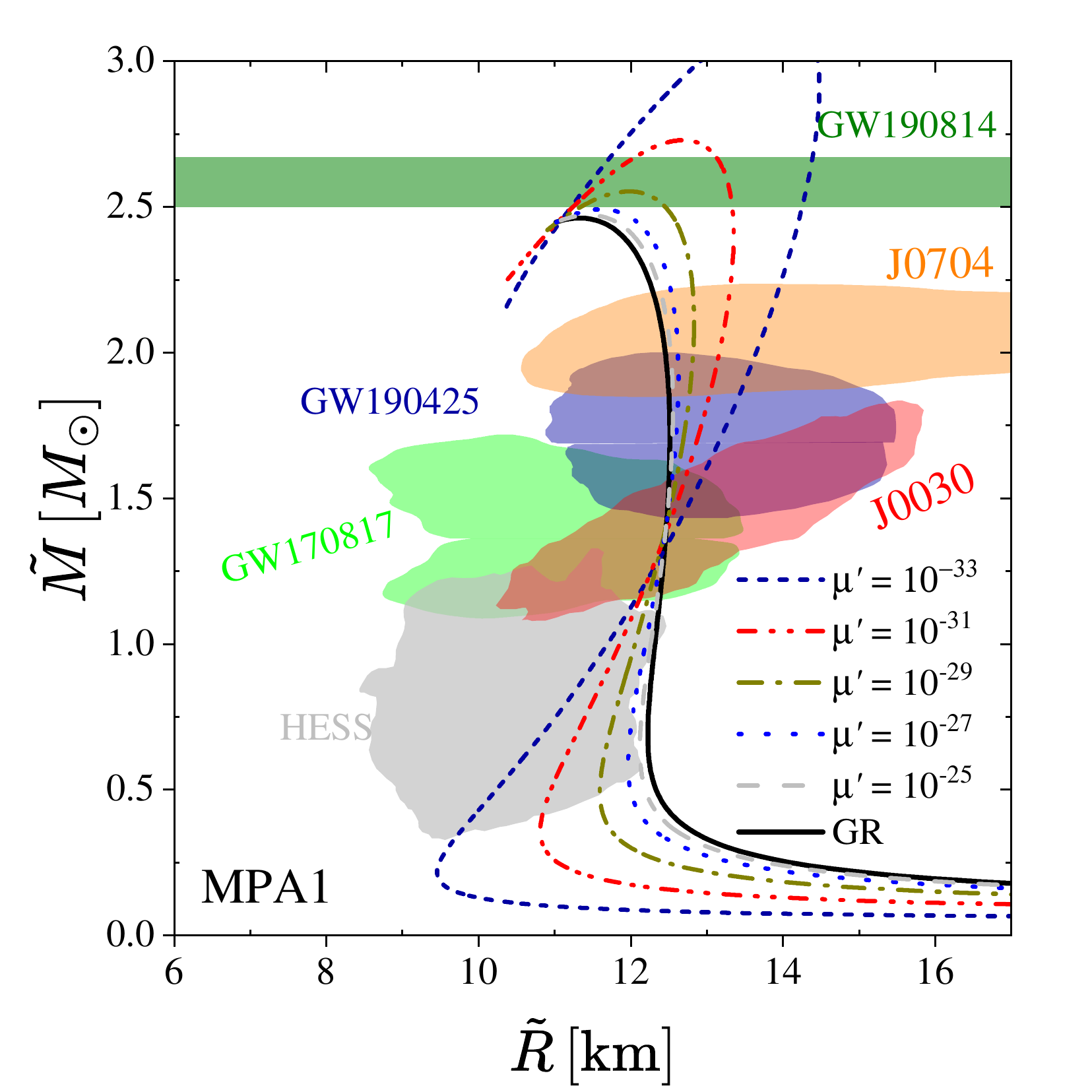}%
    \caption{ Mass-radius diagram for neutron stars for different realistic equations of state with $\gamma=6/7$, $\beta_1'=0.024$ for different values of $\mu'$. The colored domains correspond to the  recent observational mass-radius  constraints by the gravitational wave events GW170817 \cite{PhysRevLett.119.161101,PhysRevLett.121.161101}, GW190425\cite{Abbott_2020}, and GW190814 \cite{Abbott_2020_1} at $90\%$ credible level,   pulsars  PSRJ0030-0451\cite{Miller_2019}, PSRJ0740-6620\cite{Miller_2021} and HESS J1731-347 object \cite{2022NatAs...6.1444D}, at $95\%$ credible level, correspondingly. }
    \label{fig:MR2}
\end{figure*}

\begin{figure*}
    \includegraphics[width=0.49\linewidth]{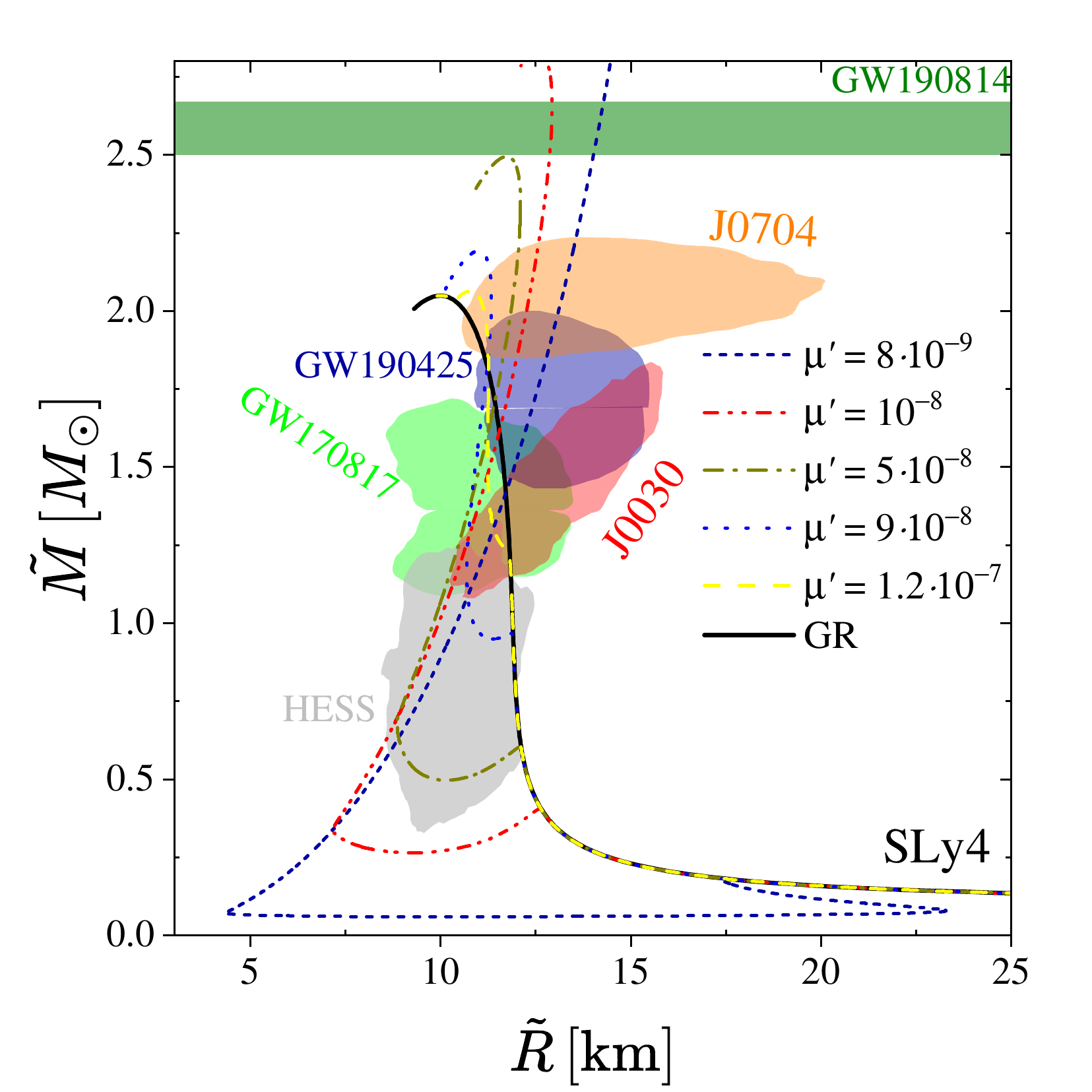}%
     \includegraphics[width=0.49\linewidth]{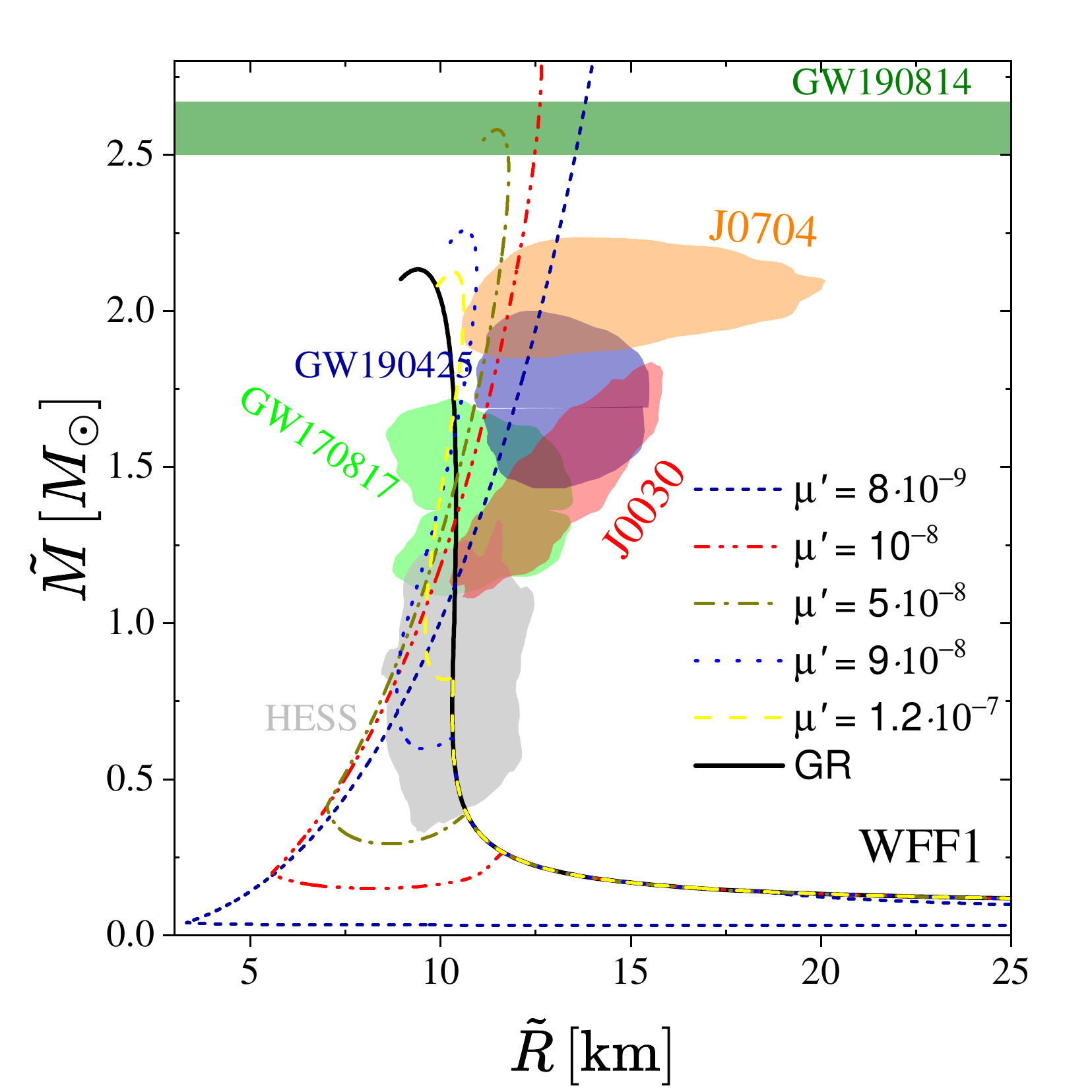}
      \includegraphics[width=0.49\linewidth]{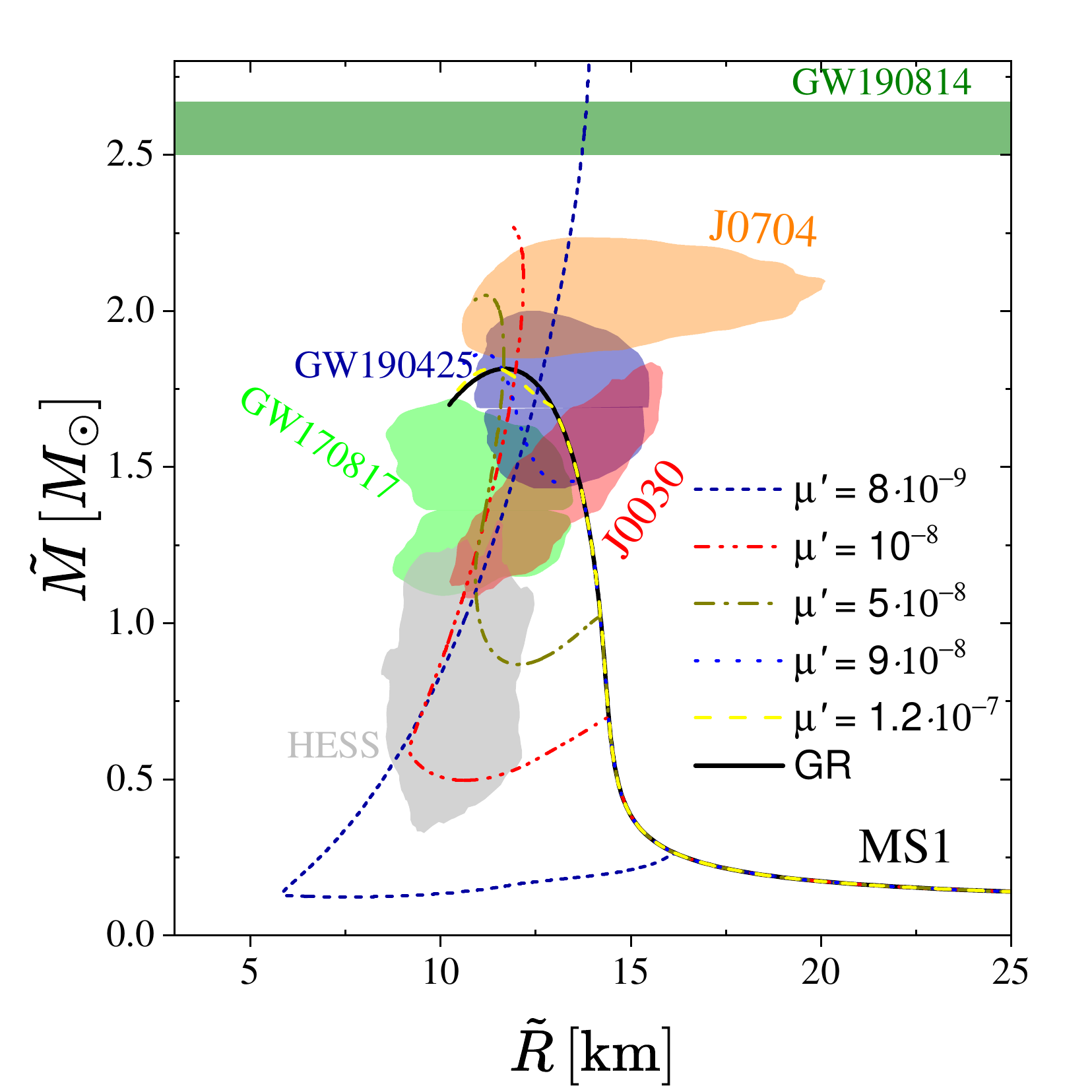}%
      \includegraphics[width=0.49\linewidth]{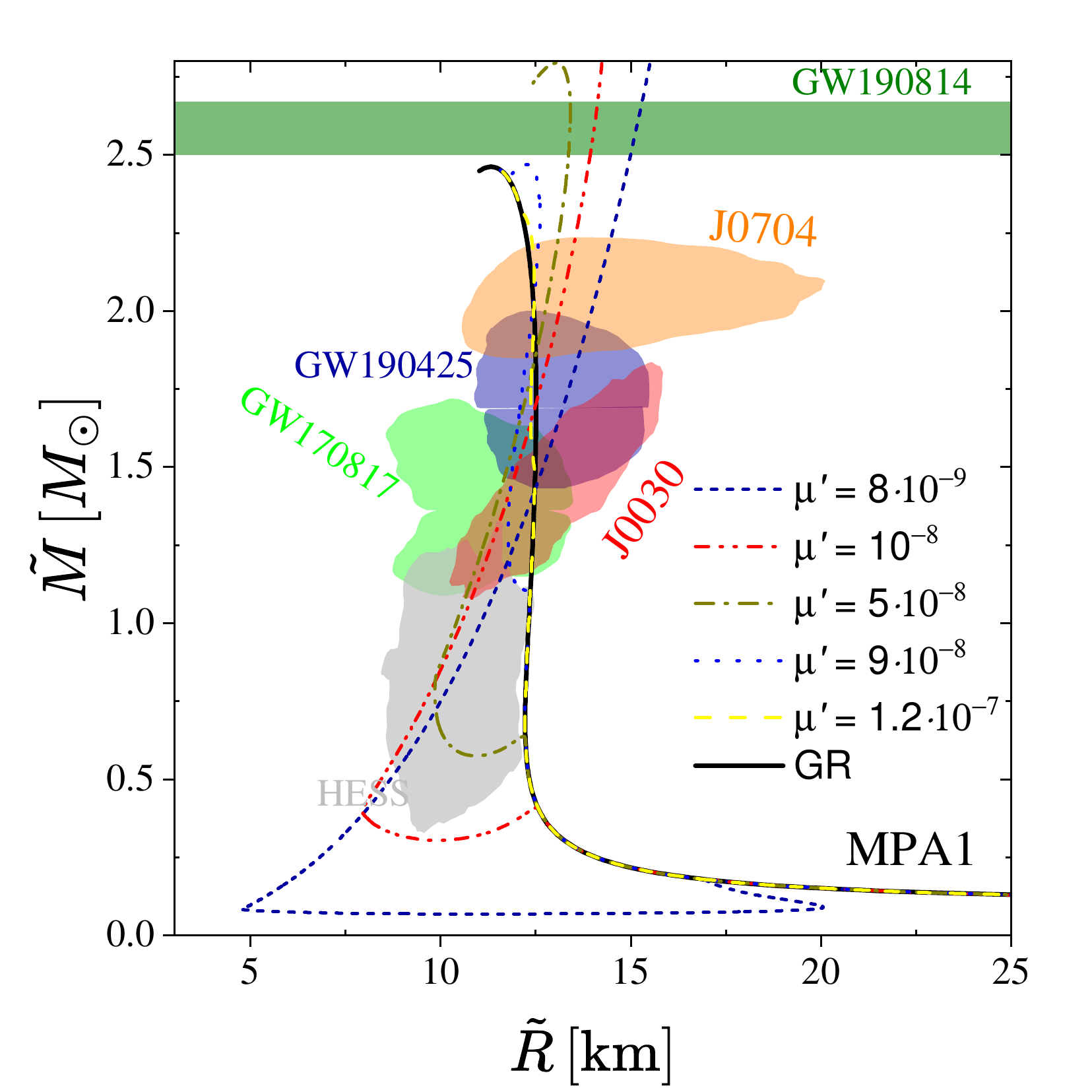}%
    \caption{ Mass-radius diagram for neutron stars for different realistic equations of state with $\gamma=0.501$, $\beta_1'=0.024$ for different values of $\mu'$. The colored domains correspond to the  recent observational mass-radius  constraints by the gravitational wave events GW170817 \cite{PhysRevLett.119.161101,PhysRevLett.121.161101}, GW190425\cite{Abbott_2020}, and GW190814 \cite{Abbott_2020_1} at $90\%$ credible level,   pulsars  PSRJ0030-0451\cite{Miller_2019}, PSRJ0740-6620\cite{Miller_2021} and HESS J1731-347 object \cite{2022NatAs...6.1444D}, at $95\%$ credible level, correspondingly. }
    \label{fig:MRp0501}
\end{figure*}

The resulting mass–radius relations for various values of $\mu'$ are shown in Figs.~\ref{fig:MR1} --\ref{fig:MRp0501} for all considered equations of state, alongside current observational constraints on the neutron star mass–radius values. These include measurements from the NICER mission for PSR J0030+0451 \cite{Miller_2019} and PSR J0740+6620 \cite{Miller_2021}, bounds from the central compact object in the supernova remnant HESS J1731–347 \cite{2022NatAs...6.1444D}, and constraints derived from the gravitational wave events GW170817 \cite{PhysRevLett.119.161101, PhysRevLett.121.161101}, GW190425 \cite{Abbott_2020}, and GW190814 \cite{Abbott_2020_1} obtained by the LIGO/Virgo–KAGRA collaborations.

We find that varying the model parameters can yield mass–radius curves consistent with observational bounds. However, decreasing $\mu'$ significantly increases both the maximum mass and the radius, pushing a large portion of the diagram beyond the observationally allowed region.

\begin{figure*}
    \includegraphics[width=0.49\linewidth]{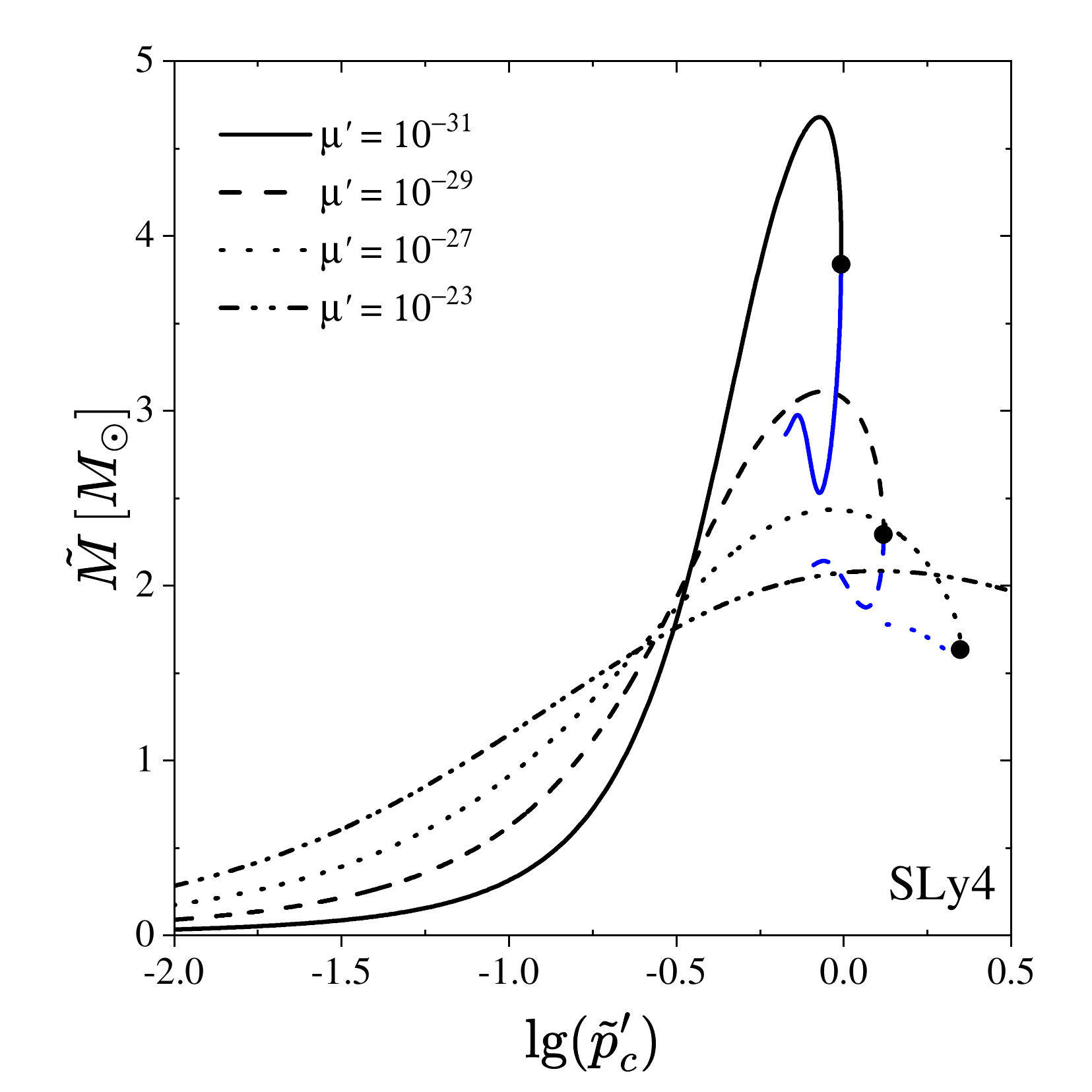}%
    \includegraphics[width=0.49\linewidth]{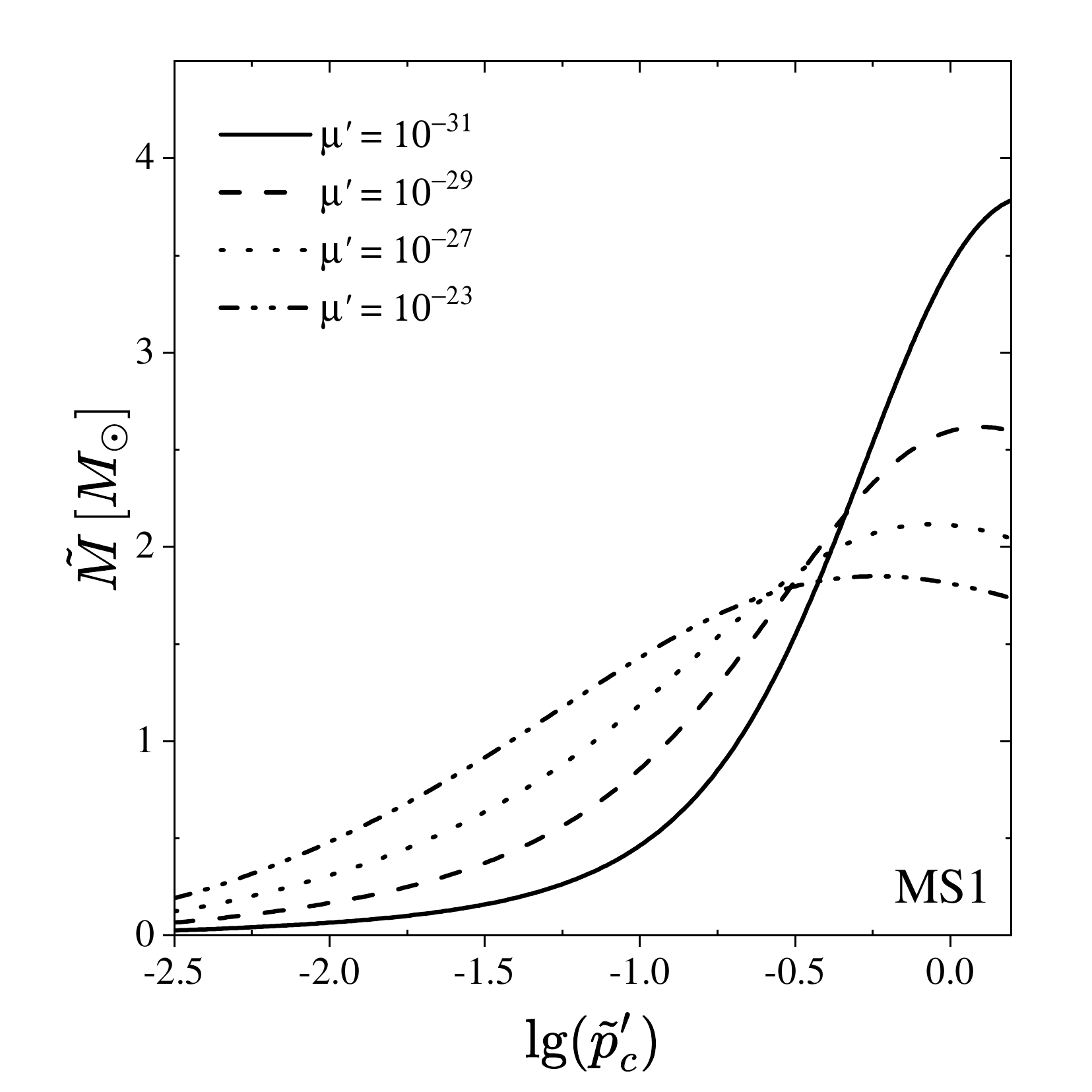}%
    
    \caption{Mass versus central pressure  for different values of $\mu'$  with $\gamma=5/6$, $\beta'=24\cdot 10^{-3}$. The black and blue lines represent the first and second branches of solutions, respectively, while the black dots indicate the maximum pressure at which these branches coincide.}
    \label{fig:Mp1}
\end{figure*}

\begin{figure*}
    \includegraphics[width=0.49\linewidth]{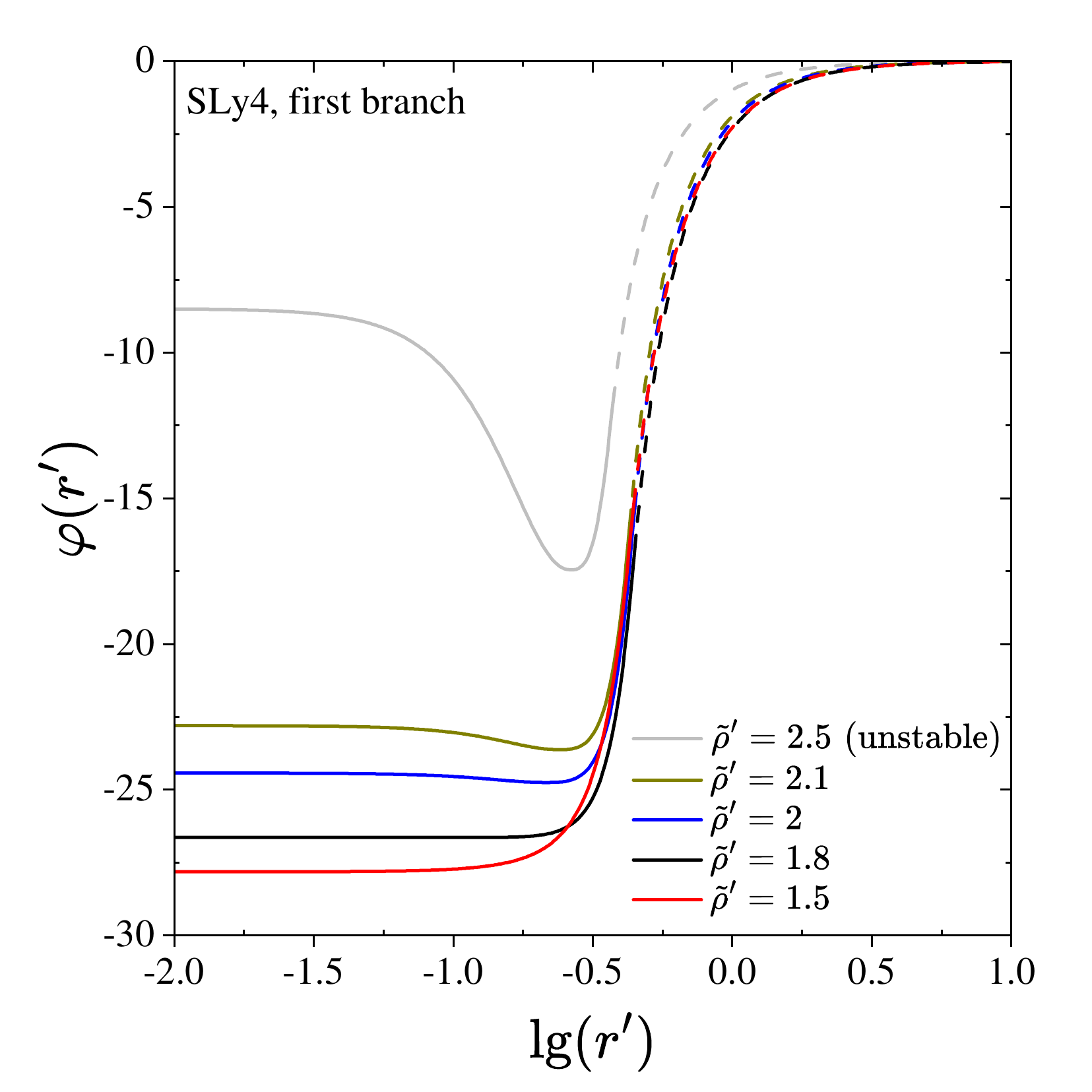}%
     \includegraphics[width=0.49\linewidth]{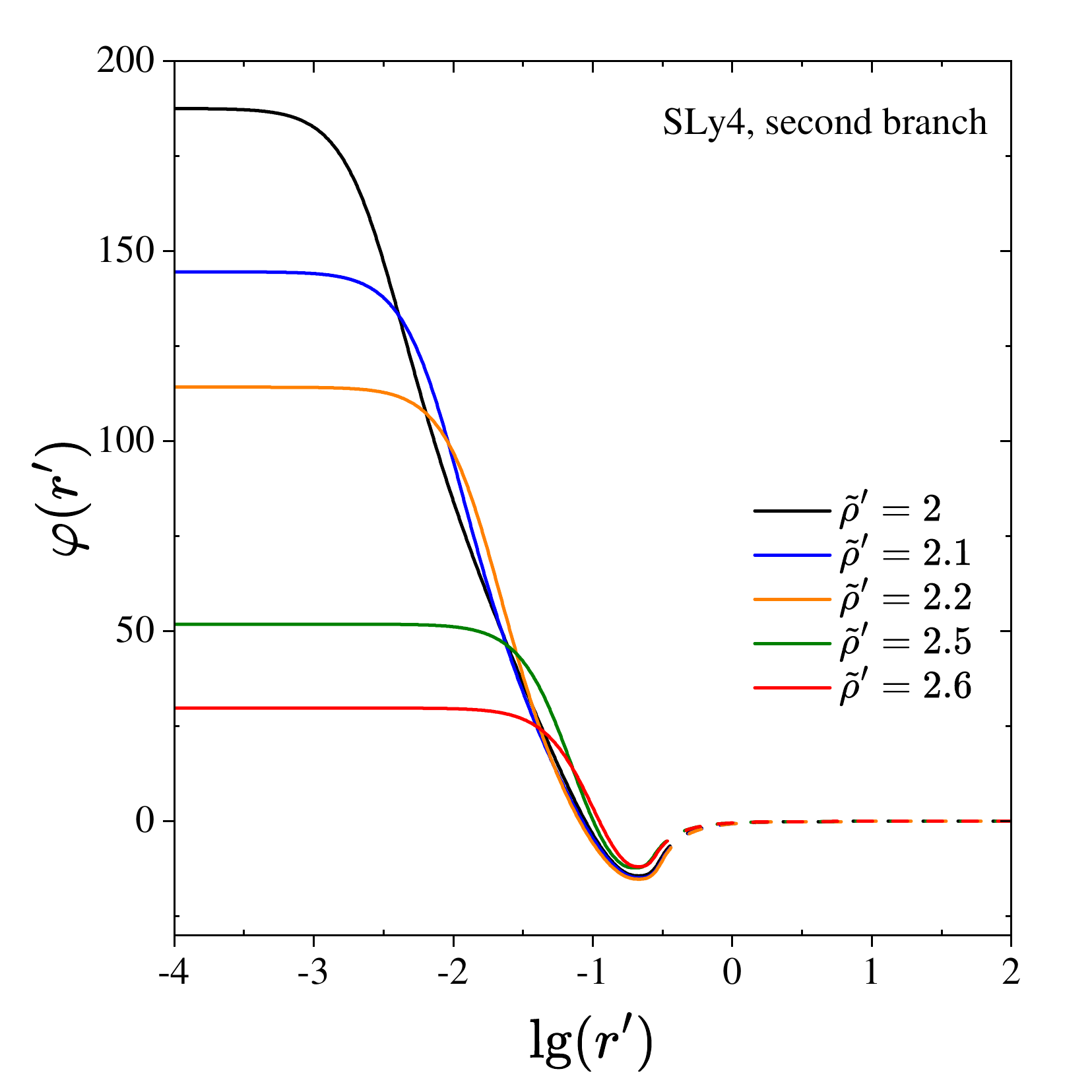}%

    \caption{The typical behavior of  scalar field  $\varphi(r)$  is shown for the first branch (left) and the second branch (right) as functions of radius $r$s, for the SLy4 EoS with $\beta_1'=0.024$, $\gamma=5/6$, and $\mu'=10^{-29}$, for several values of the central density. The solid and dashed segments correspond to the inner and outer parts of the solution, respectively.
}
    \label{fig:scalar_field}
\end{figure*}

\begin{figure*}
    \includegraphics[width=0.49\linewidth]{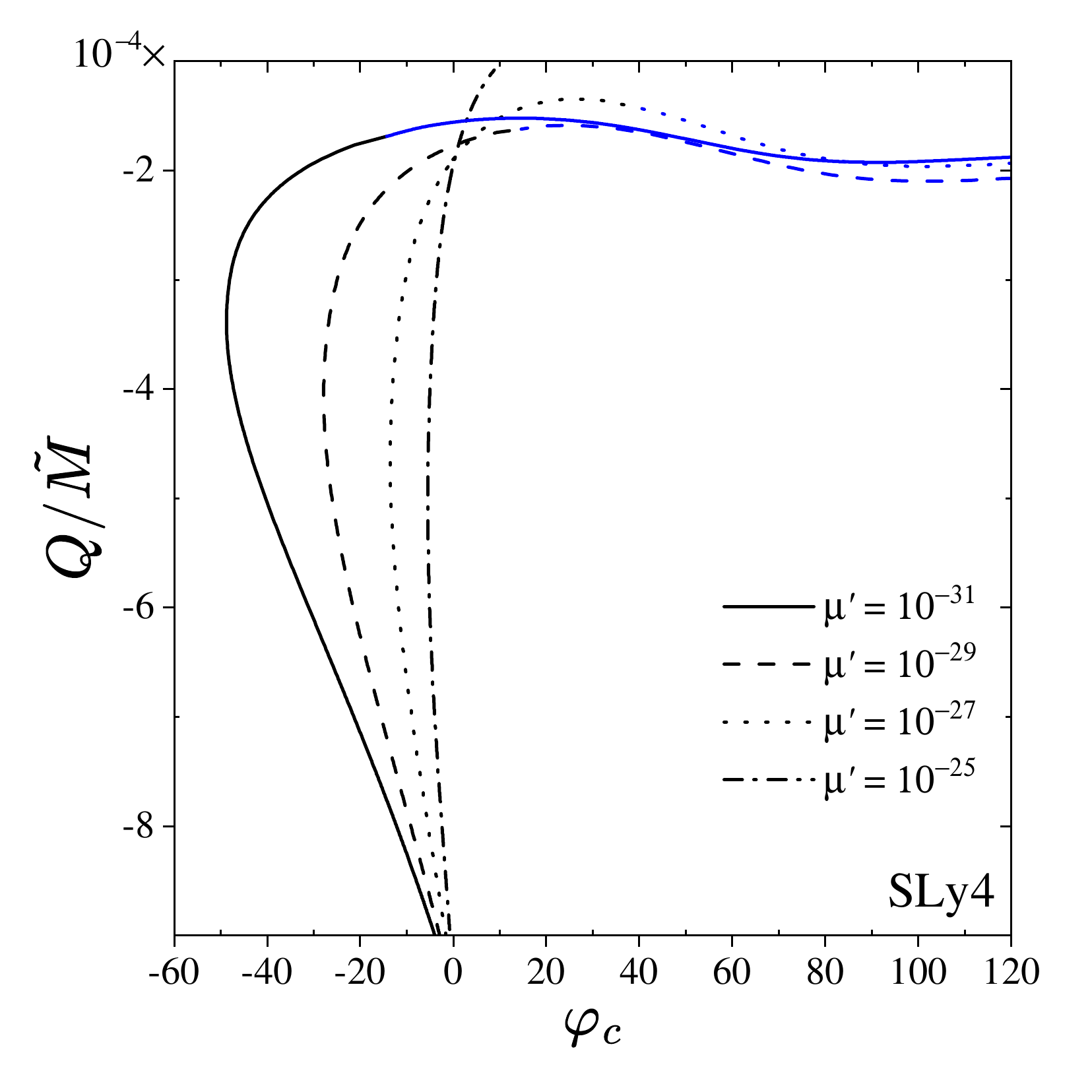}%
    \includegraphics[width=\columnwidth]{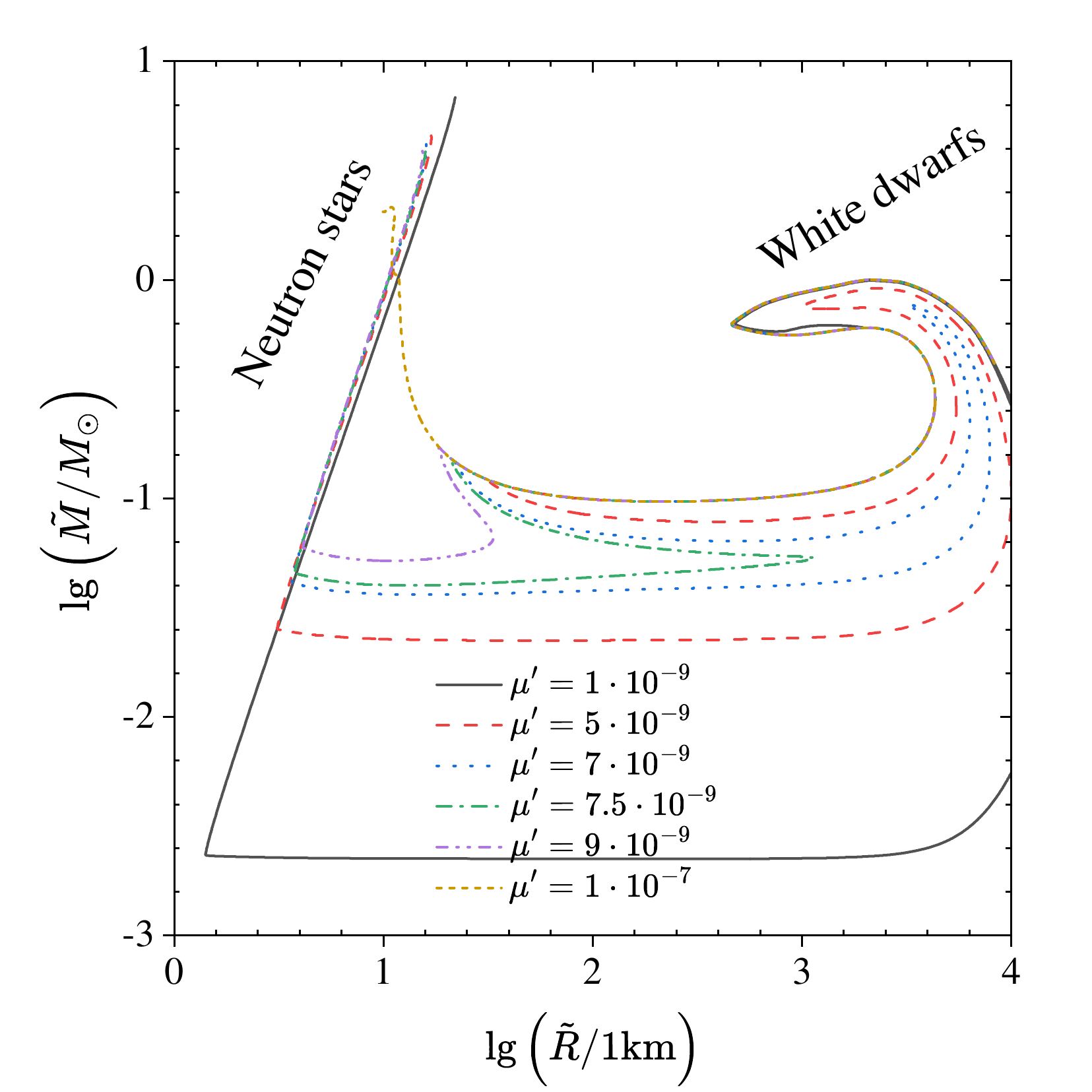}%
    
    \caption{Left panel: The typical behavior of the rescaled scalar charge $Q/\tilde{M}$ as a function of $\varphi_c$ for the SLy4 EoS with $\beta_1' = 0.024$ and $\gamma = 5/6$. Right panel: Detailed behavior of the MR curves in log scale for $\gamma = 0.501$ at different values of $\mu$.}
    \label{fig:scalar_charge}
\end{figure*}

\begin{figure*}
    \includegraphics[width=0.49\linewidth]{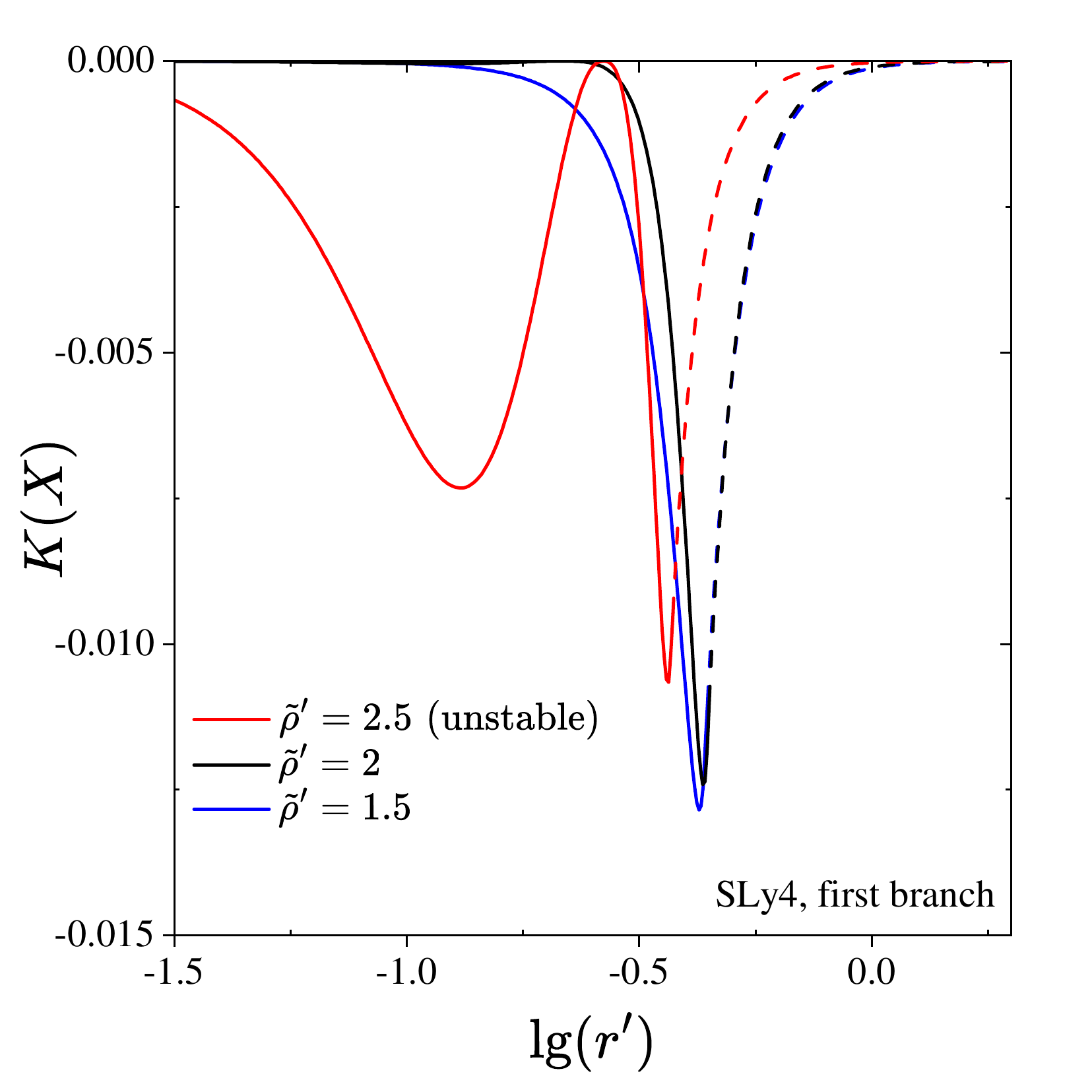}%
     \includegraphics[width=0.49\linewidth]{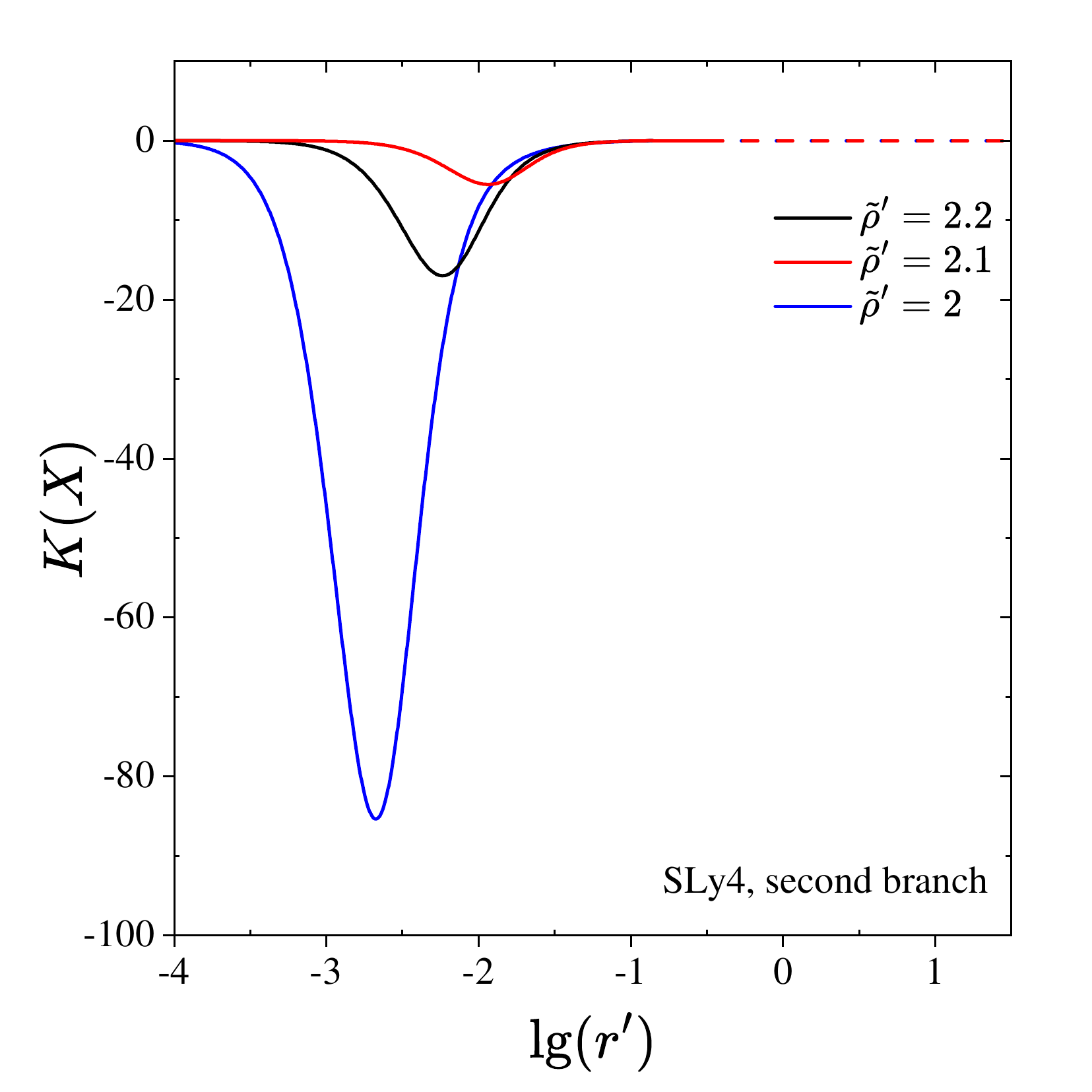}%
     
    \caption{The typical behavior of the rescaled kinetic  term $K(r)$ is shown for the first branch (left) and the second branch (right) as functions of radius, for the SLy4 EoS with $\beta_1' = 0.024$, $\gamma = 5/6$, and $\mu = 10^{-29}$, for several values of the central density. The solid and dashed segments correspond to the inner and outer parts of the solution, respectively.
}
    \label{fig:K-essence}
\end{figure*}

Additionally, a second branch of solutions appears, for a given central density $\tilde{\rho}_c$, two distinct central scalar field values $(\varphi_c^{(1)}, \varphi_c^{(2)})$ can lead to configurations satisfying the correct asymptotic behavior of $\varphi(r)$ at spatial infinity; see Fig.~\ref{fig:Mp1}, left. The typical profiles of $\varphi(r)$ and the normalized scalar charge for both branches are shown in Figs.~\ref{fig:scalar_field} and \ref{fig:scalar_charge}, respectively, while the corresponding behavior of the $K$-essence function $K(r)$ is presented in Fig.~\ref{fig:K-essence}. Along the first branch, $\varphi(r)$ is nearly monotonic. In contrast, on the second branch, the scalar field exhibits non-monotonic behavior and changes sign.

The second branch is characterized by large values of $\varphi_c \gg 0$ at relatively small $\tilde{\rho}_c$, which renders the numerical integration more challenging. At a critical value $\tilde{\rho}_c = \tilde{\rho}_c^{\text{ext}}$, the two branches merge. For $\tilde{\rho}_c > \tilde{\rho}_c^{\text{ext}}$, we are unable to find solutions that exhibit acceptable asymptotic scalar field behavior.

We note that the second branch predominantly connects to the unstable segment of the first branch, suggesting that it may be dynamically unstable. This branch also lies in the region corresponding to the unstable solutions in general relativity (GR), defined by the segment beyond the first sign change in $\partial \tilde{M} / \partial \tilde{\rho}_c$. However, increasing $\mu'$ shifts the merging point toward the region associated with potentially stable configurations. As $\mu'$ increases, the mass–radius diagram approaches the GR limit.

For $\gamma = 1/2+\epsilon$ (small $\epsilon$), where anti-screening effects are still present, the $M$–$R$ diagram develops additional structural features (see Fig.~\ref{fig:MRp0501} and Fig.~\ref{fig:scalar_charge}, right panel) in the intermediate interval $\mu' \in (\mu'_1,\mu'_2)$. In these figures we have $\epsilon=0.001$.  For small values $\mu' < \mu'_1$ the behavior remains similar to the previous case. For $\mu' > \mu'_1$, the loop-like branch of the $M$–$R$ curve that initially corresponds to the white-dwarf branch ($\rho_c \lesssim 10^{11},\mathrm{g/cm^{3}}$) begins to deform at a turning point and gradually transforms into an extended, nose-like structure with changing $\mu'$ (see Fig.~\ref{fig:scalar_charge}, right panel). At the same time, the inner loop starts to deform by  migrating toward lower masses and radii, finally  forming an S-like shape. As $\mu'$ changes, this behavior occurs at  higher central densities, in some cases reaching values characteristic of the liquid core of neutron stars ($\rho_c \gtrsim 10^{14},\mathrm{g/cm^{3}}$). For $\mu' > \mu'_2$ the curve smoothens and approaches the GR limit. Most of the S-like segments satisfy $\partial \tilde{M}/\partial \rho_c < 0$, indicating instability, whereas isolated portions with positive slope might  correspond to stable configurations, analogous to the third family of compact stars \cite{1983bhwd.book.....S}.

\section{Conclusions}
In this paper, we studied the impact of a novel class of gravitational models with a scalar that carries a non-canonical kinetic term. These models trivially obey solar system bounds by having a small coupling to matter. But become important in the regime of very high densities. We examined their impact on static, spherically symmetric neutron stars with realistic equations of state. A key feature of the considered model is the anti-screening of the scalar-mediated fifth force in high-density environments, such as the interiors of neutron stars.

We demonstrated that the model parameters have a significant impact on neutron star structure. For suitable parameter choices, it  can successfully reproduce current observational constraints on neutron star masses and radii. Additionally, we can obtain configurations with maximum masses significantly exceeding $3M_\odot$. Furthermore, we found that for a fixed central pressure (or central density), the system can admit two distinct configurations, corresponding to different central values of the scalar field and different scalar charges at spatial infinity. These configurations form two separate branches of solutions. We demonstrated the presence of a critical value of central pressure at which these branches merge, and above which no solutions with acceptable asymptotic behavior can be found.
\label{sec:conclusions}

\acknowledgments
M.~P.~H. is supported in part by National Science
Foundation grant PHY-2310572.
O.~S.~S. is supported in part by National Science 
Foundation grant PHY-2110466 and the Tufts Scholar at Risk Program.

\FloatBarrier
\bibliography{references.bib}

\end{document}